\documentclass[journal]{IEEEtran}

\usepackage{epstopdf}
\usepackage{graphicx}
\usepackage{stmaryrd}
\usepackage{amsmath,amsthm,amssymb}
\usepackage{multirow,url}
\usepackage{color,cite}
\usepackage[ruled,vlined,linesnumbered]{algorithm2e}

\usepackage{setspace}
\usepackage{rotating}

\DeclareMathAlphabet{\mathpzc}{OT1}{pzc}{m}{it}

\newcommand{\m}{\operatornamewithlimits{m}}

\newcommand{\poa}{\operatornamewithlimits{PoA}}
\newcommand{\Ne}{\operatornamewithlimits{Ne}}
\newcommand{\mbps}{\operatornamewithlimits{Mbps}}

\newtheorem{lemma}{Lemma}

\newtheorem{definition}{Definition}
\newtheorem{theorem}{Theorem}

\ifodd 0
\newcommand{\rev}[1]{{\color{black}#1}} 
\newcommand{\com}[1]{\textbf{\color{white} (COMMENT: #1)}} 
\newcommand{\comg}[1]{\textbf{\color{white} (COMMENT: #1)}}
\newcommand{\response}[1]{\textbf{\color{white} (RESPONSE: #1)}} 
\newcommand{\edi}[1]{{\color{green}#1}}
\else
\newcommand{\rev}[1]{#1}
\newcommand{\com}[1]{}
\newcommand{\comg}[1]{}
\newcommand{\response}[1]{}
\newcommand{\edi}[1]{#1}
\fi

\begin{document}

\title{Quality of Service Games for Spectrum Sharing}

\author{Richard Southwell, Xu Chen, \emph{Member, IEEE}, and Jianwei Huang, \emph{Senior Member, IEEE} \thanks{
Richard Southwell and Jianwei Huang are with the Network Communications and Economics Lab, Department of Information Engineering, the Chinese University of Hong Kong (emails:richardsouthwell254@gmail.com, jwhuang@ie.cuhk.edu.hk).

Xu Chen (corresponding author) is with the School of Electrical, Computer and Energy Engineering, Arizona State University, Tempe, Arizona, USA (email:xchen179@asu.edu).
}

}

\maketitle
\pagestyle{empty}
\thispagestyle{empty}

\begin{abstract}
Today's wireless networks \rev{are increasingly crowded with an explosion of wireless users, who have} greater and more diverse quality of service (QoS) demands than ever before. However, \rev{the amount of} spectrum that can be used to satisfy these demands \rev{remains finite. This leads to a great challenge for wireless users to effectively share the spectrum to achieve their QoS requirements}. This paper presents a game theoretic model for spectrum sharing, where users seek to satisfy their QoS demands in a distributed fashion. \rev{Our spectrum sharing model is quite general,} because \rev{we allow} different \rev{wireless} channels \rev{to} provide different QoS,  depending upon their channel conditions and how many users are trying to access them.
Also, users can be highly heterogeneous, with different QoS demands, depending upon their activities, hardware \rev{capabilities}, and technology choices. Under such a general setting, we show that it is NP hard to find a spectrum allocation which satisfies the maximum number of users' QoS requirements in a centralized fashion.  We also show that allowing users to self-organize through distributed channel selections is a viable alternative to the centralized optimization, because better response updating is guaranteed to reach a pure Nash equilibria in polynomial time. By bounding the price of anarchy, we demonstrate that the worst case pure Nash equilibrium can be close to optimal\rev{, when users and channels are not very heterogenous}. We also extend our model by considering the frequency spatial reuse, and consider the user interactions as a game upon a graph where players only contend with their neighbors. \rev{We prove that better response updating is still guaranteed to reach a pure Nash equilibrium in this more general spatial QoS satisfaction game.}
\end{abstract}


\section{Introduction}\label{intro}

The number of wireless devices such as smart-phones continues to increase rapidly in today's market, while the amount of spectrum available for these devices remains limited. Moreover, many new wireless applications such as high definition video streaming and online interactive gaming are emerging, making the quality of service (QoS) demands of wireless users \rev{higher} and more varied. Thus there is an urgent need to study the issue of how to efficiently \rev{share} the limited spectrum to satisfy the QoS demands of as many users as possible.

There are two different approaches to address this issue. The first approach is a \emph{centralized} approach, where a network operator optimizes the spectrum \rev{resources} to meet the users' QoS requirements.  This approach puts most of the implementation complexity at the operator side, and wireless devices do not need to be very sophisticated. However, as the networks grow larger and more heterogeneous, this approach can become unsuitable for two reasons. Firstly, the QoS demands of wireless users are highly heterogeneous, which implies that the operator needs to gather massive amounts of information from users in order to perform the centralized optimization. Secondly, finding the system-wide optimal QoS demand satisfaction solution is computationally challenging -- in fact we show that it is NP hard. It is hence difficult for the operator to compute the optimal solution to meet users' real-time QoS demands. The alternative approach is a \emph{decentralized} approach, where each wireless user makes the spectrum access decision locally to meet its own QoS demand, while taking the network dynamics and other users' actions into consideration. This is feasible since new technologies like cognitive radio \cite{liang} give users the ability to scan and switch channels easily. The decentralized approach enables more flexible spectrum sharing, scales well with the network size, and is particular suitable when users belong to multiple network entities.

In this paper, we focus on the decentralized approach, and propose a new framework of \emph{QoS satisfaction games} to model the distributed QoS demand satisfaction problem among the users. Game theory is a useful tool for designing distributed algorithms that allow users to self-organize, optimize their channel selections, and satisfy their QoS demands. Our QoS satisfaction game framework is developed, based on the theory of congestion games \cite{rosenthal}. The central idea behind congestion games is that there are many \emph{players}, each of which selects a \emph{resource} to use. A player's  utility is a non-increasing function of the total number of players using the same resource. The distributed QoS satisfaction problem can be modeled using congestion games by thinking of the players as wireless users, while the resources represent different channels \cite{mingyan}. The satisfaction of a user's QoS demand depends on its congestion level, i.e., how many users are competing for its channel. \edi{In our QoS satisfaction game, a player achieves  a unit utility when its  channel's data rate is sufficiently high to satisfy its QoS demand. Otherwise, the player's utility is negative and it is better off by switching channels (to improve the payoff)  or turning off its transmitter (to receive a zero payoff).}

\subsection{Related Work}\label{related}

Rosenthal proposed the original congestion game model \cite{rosenthal} for the scenario where different resources can have different utility functions associated with them (i.e., heterogenous resources) but all players have the same utility function for any particular resource (i.e., homogenous players). This kind of system has a pleasing feature known as the \emph{finite improvement property} - which means that when the system evolves because players asynchronously perform better response updates (i.e., the players selfishly improve their resource choices), the system is guaranteed to reach a pure Nash equilibrium in a finite number of steps. A \emph{pure Nash equilibrium} is a system state where no player has any incentive to deviate unilaterally.

However, the original congestion game is not general enough to model spectrum sharing, because it assumes that players are homogenous, whereas wireless network users are often highly heterogenous. The congestion games with player-specific utility functions considered in \cite{playerspec} are more appropriate for this modeling purpose. Authors in \cite{mingyan,Law2009,xu2011,specmob} have adopted such a game model for studying spectrum sharing problems. However, unlike classical congestion games, these games are not necessarily guaranteed to possess the finite improvement property.

\emph{Spatial reuse} is another feature of wireless networks that the original congestion game model does not account for. In reality only \emph{nearby} users on the same channel will interfere with each other. Users which are distantly separated will not cause congestion to each other. A congestion game on a graph can be used to realistically capture the spatial aspect of spectrum sharing. The idea behind such a system is that a user's utility only depends upon the number of users of the same channel \emph{who are linked to them in the graph}. In \cite{cem}, we introduced a general class of congestion games on graphs that are appropriate for modeling \rev{spectrum sharing}. Although there are many subclasses of these games which always admit the finite improvement property, we demonstrated that there exist congestion games on graphs that  do not have \emph{any} pure Nash equilibria. We have also further developed several more elaborate  graphical congestion game models \cite{gamenets11,gamenets12,arxiv,SpatialXu} with applications to \rev{spectrum sharing}.

A common assumption within most previous congestion game based spectrum sharing literature (e.g., \cite{mingyan,Law2009,xu2011,specmob,gamenets11,SpatialXu}) is that a user's utility strictly increases with its received data rate (and hence strictly decreases with the congestion level). This is true, for example,  when  users are running elastic applications such as file downloading. However, there are many other types of applications with more specific QoS requirements, such as VoIP and video streaming. These inelastic  applications cannot work properly when their QoS requirements (e.g., data rates) are unmet, and do not \rev{enjoy} any additional benefits when given more resources than needed. This kind of traffic is becoming increasingly popular over the wireless networks (e.g., mobile video traffic exceeded $50\%$ percent of all wireless traffic in 2011 according to the report by Cicso \cite{cicso}). \emph{This motivates the QoS satisfaction game model in this paper.}

Rather than assuming that users wish to increase their data rates whenever possible, we assume that each user has a fixed QoS demand. \rev{If} the demand is  satisfied, then the user has no inclination to change his choice of resource. Our game model was inspired by the \emph{games in satisfaction form} considered in \cite{satform}. In \cite{satform} the authors \rev{considered} other games where players wish to satisfy demands, and the authors design algorithms to find \emph{satisfaction equilibria}, which are strategy profiles where all users are satisfied. In our paper, we consider the more general case where some users' QoS requirements may not be satisfied (given the limited spectrum resource). The case where a satisfaction equilibrium exists becomes a special case of our model. Moreover, we also take \rev{into} account the issue of spatial reuse. This makes the modeling more practical for wireless communication systems. The generalizations considered in our model result in more challenges and significant differences in analysis.

\edi{When discussing the achievability of the equilibrium, we focus on dynamics where one player can perform a better response update each time. There are many alternative types of dynamics we could consider, such as smoothed best response dynamics and imitation dynamics \cite{chen2012imitative}. We could also consider the replicator dynamics from evolutionary game theory. Reference \cite{Chen2013TMC} showed that replicator dynamics can be used for  spectrum sharing using appropriate message passing protocols. Replicator dynamics is most useful  when the user population is large, and in which case the system will follow  continuous (essentially deterministic) dynamics which normally converge to evolutionarily stable strategies. However, many techniques from evolutionary game theory rely upon the assumption the players are homogenous, while our wireless users are typically heterogenous. Another issue is that translating replicator dynamics into the spatial setting (i.e., a game on a graph) is quite difficult.}

\subsection{Contributions}\label{contribution}


Our main results and contributions can be summarized as follows:

\begin{itemize}
\item \emph{A general QoS satisfaction game framework:} We formulate the distributed QoS demand satisfaction problem among wireless users as a QoS satisfaction game, which is general enough to capture the details of spectrum sharing over a wide range of scenarios, with heterogenous channels and users. \edi{Despite allowing for heterogenous channels, heterogenous users, and spatial interactions, we still obtain several significant analytic results}.

\item \emph{Remarkable convergence properties:} We prove that every QoS satisfaction game has the finite improvement property. This is remarkable because many congestion games with heterogenous resources and players do not have this feature. More importantly, it enables us to design a distributed QoS satisfaction algorithm which allows wireless users to easily self-organize into a pure Nash equilibrium.

\item \emph{Spatial generalization:} We generalize the model by thinking of users as vertices, which are linked in a graph and can only interfere with their neighbors. We show that the resulting QoS games on graphs also possess the finite improvement property.
\end{itemize}



The rest of the paper is organized as follows. We introduce the QoS satisfaction game model and study its properties in Sections \ref{model} and \ref{results}, respectively. We then generalize the game model with spatial reuse in Section \ref{spatial}. We then propose the distributed QoS satisfaction algorithm and evaluate its performance by simulations in Section \ref{dist}. Finally, we conclude the paper in Section  \ref{conc}. \textbf{Most proofs are provided in the Appendix.}

\section{QoS Satisfaction Game}\label{model}

In this section we formally define the QoS satisfaction game model for spectrum sharing. Spectrum sharing is a promising approach to address the spectrum under-utilization problem. Field measurements by Shared Spectrum Cooperation in Chicago area shows that the overall average utilization of a wide range of different types of spectrum bands is lower than $20\%$ \cite{SOM2005}. In order to improve the overall spectrum utilization, several countries have recently reformed their policy (such as the FCC's ruling for the TV white space \cite{FCC}) and allow spectrum sharing, such that unlicensed users equipped with cognitive radios can access the channels which are tentatively not used by the licensed spectrum users. In this paper, we consider the spectrum sharing problem among multiple unlicensed users who run different applications and hence have heterogeneous QoS demands.

\subsection{Game model}\label{physics}

A \textbf{QoS satisfaction game} is defined by a tuple $(\mathcal{N}, \mathcal{C}, (Q_n^c)_{n \in \mathcal{N}, c \in \mathcal{C}}, (D_n)_{n \in \mathcal{N}})$ where:
\begin{itemize}
\item $\mathcal{N}=\{1,\edi{\ldots},N\}$ is the set of \textbf{wireless unlicensed users}, also referred as the \textbf{players}.
\item $\mathcal{C}=\{1,\edi{\ldots},C\}$ is the set of \textbf{channels}. Each unlicensed user may select one channel to access. Furthermore, we introduce the element $0$ to represent the \textbf{dormant state}. Choosing the dormant state will be beneficial when an unlicensed user's QoS demand cannot be satisfied due to limited resources. In such a case the user can choose the dormant state $0,$ which corresponds to ceasing its transmission to save power consumption. Now we use the term `dormant state' instead of `virtual channel' since it involves introducing less new concepts, and we no longer have to speak of ``real'' channels. In summary, each unlicensed user/player has a strategy set $\tilde{\mathcal{C}}=\{0,1,\edi{\ldots},C\}$ which consists of all channels, together with the dormant state. The \textbf{strategy profile} of the game is given as $\boldsymbol{x} = (x_1,x_2,\edi{\ldots},x_N) \in \tilde{\mathcal{C}}^N,$ where each unlicensed user $n$ chooses a strategy $x_n \in \tilde{\mathcal{C}}$.

\item $Q_n^c(\cdot)$ is a non-increasing function that characterizes \rev{the} data rate received by an unlicensed user $n$ \rev{who} has selected channel $c$. Specifically, we have $Q_n^c(I^c(\boldsymbol{x}))=\theta_n^c B_n^c g_n^c(I^c(\boldsymbol{x}))$, with $I^c(\boldsymbol{x}) = |\{n \in \mathcal{N}: x_n = c \}|$ being the \textbf{congestion level} of channel $c$, i.e., the number of users who choose channel $c$. We detail the parameters in $Q_n^c$ as follows.
    \begin{itemize}
    \item $\theta_n^c\in\{0,1\}$ is the channel availability indicator.  When channel $c$ is occupied by licensed users and not available for unlicensed user $n$, we have $\theta_n^c=0$, in which case $Q_n^c(I^c(\boldsymbol{x}))=0$ \rev{for any value of $I^c(\boldsymbol{x})$}. For a limited period of time, the usage of spectrum by licensed users is assumed to be static (but can change in different periods)\footnote{We show in Theorem \ref{graphical} that the proposed QoS satisfaction game algorithm can converge in a fast manner (e.g., less than one second in practical 802.11 systems). In this case, as long as the activities of licensed users change in a larger timescale in terms of seconds/miniutes/hours (e.g., TV/daytime radio broadcasting), we can still implement the QoS satisfaction game solution for the system.}. This is appropriate for modeling the TV spectrum, for example, where the activities of licensed users change very slowly. According to the most recent ruling by the FCC, unlicensed users can \rev{reasonably and  accurately} determine the spectrum availability \rev{within a short amount of time by consulting a} database \cite{FCC}.  When channel $c$ is available for the spectrum access by an unlicensed user $n$ (i.e., $\theta_n^c=1$), we have $Q_n^c(I^c(\boldsymbol{x}))>0$.

    \item $B_n^c$ is the mean channel throughput of user $n$ on channel $c$. We allow user specific throughput functions, i.e., different users may have different $B_n^c$ even on the same channel $c$. This enables us to model users with different transmission technologies, different coding/modulation schemes, different channel conditions\rev{, and different reactions to the same licensed user on the channel}. For example,  we can compute the maximum channel throughput $B_{c}^{n}$ according to the Shannon capacity as\begin{equation}
    B_{c}^{n}=W_{c}\log_{2}\left(1+\frac{\zeta_{n}z_{c}^{n}}{\omega_{c}^{n}}\right),\label{eq:dd}\end{equation}
    where $W_{c}$ is the bandwidth of channel $c$, $\zeta_{n}$ is
    the fixed transmission power adopted by user $n$ according to the requirements such as the primary user protection, $\omega_{c}^{n}$ denotes
    the background noise power, and $z_{c}^{n}$ is the user-specific channel gain.

    \item $g_n^c(I^c(\boldsymbol{x}))$ is the channel contention function that describes the probability that user $n$ can successfully grab the channel $c$ for data transmissions given the congestion level $I^c(\boldsymbol{x})$. In general, $g_n^c(I^c(\boldsymbol{x}))$ decreases as the number of contending users $I^c(\boldsymbol{x})$ increases. For example,  if we adopt the TDMA mechanism for the medium access control (MAC) to schedule users in the round-robin manner, then we have $g_n^c(I^c(\boldsymbol{x}))=\frac{1}{I^c(\boldsymbol{x})}$.
    \end{itemize}
\item $D_n \geq 0$ is the data rate demand of unlicensed user $n$. For example, listening to an MP3 online will require a small  $D_n$, whereas watching a high definition streaming video requires a large $D_n$.
\end{itemize}

The \textbf{utility} of an unlicensed user $n$ in strategy profile $\boldsymbol{x}$ is
\begin{equation} \label{uo}
  U_n(\boldsymbol{x})=\begin{cases}
    1, & \text{if $x_n \neq 0$ and $Q_n^{x_n}(I^{x_n}({\boldsymbol x})) \geq D_n$},\\
    0, & \text{if $x_n = 0$},\\
    -1, & \text{if $x_n \neq 0$ and $Q_n^{x_n}(I^{x_n}({\boldsymbol x})) < D_n$}.\\
  \end{cases}
\end{equation}
A \textbf{satisfied user} is an unlicensed user $n$ \rev{who chooses} a channel $x_n\neq 0$ \rev{and receives a} data rate $Q_n^{x_n}(I^{x_n}({\boldsymbol x}))$ \rev{not smaller than} its QoS demand $D_n$. A satisfied users receives a utility of $1$. A \textbf{dormant user} is an unlicensed user $n$ choosing the dormant state $x_n = 0$. Such a dormant user \rev{does not receive any benefit (as it achieves a zero data rate) or any penalty (as it does not waste any energy)}, and gets a utility of $U_n (\boldsymbol{x})= 0$. A \textbf{suffering user} is an unlicensed user $n$ \rev{who chooses} a channel $x_n\neq 0$ \rev{but receives a} data rate $Q_n^{x_n}(I^{x_n}({\boldsymbol x}))$ below its QoS demand $D_n$. Such a suffering user expends power without gaining \rev{any} benefit, and so it gets a utility of $U_n (\boldsymbol{x})= -1$.

A suffering user can always increase their utility by becoming dormant without harming any other user. This suggests that rational \rev{(i.e., utility maximizing)} players will \rev{eventually end up at} strategy profiles which contain no suffering users. We say that a strategy profile is \textbf{natural} if it holds no suffering users.

It is worth noting that we can easily generalize our model by allowing an unlicensed user $n$ to receive a utility of $u_n$ if it is satisfied, $v_n$ if it is  dormant, and $t_n$ if it is suffering, where $u_n > v_n >t_n$. Making this generalization does not \edi{affect} the better response dynamics or the set of pure Nash equilibria discussed later on, because the preference orderings of the strategies in the generalized game are the same as in our current model\footnote{Technically speaking, our game is weakly isomorphic \cite{weakiso} to this generalized version.}. Our results about convergence (Theorem \ref{completeconvergence}) and computational complexity (Theorem \ref{nphard}) also remain true for games with generalized utility functions. However, since the generalized games allow different players to receive different utilities when satisfied, our results  about social optimality (Theorems \ref{homplayequiv} and \ref{algodworks}) may not hold for the generalized games. In this paper, we will restrict our attention to the utility choices of $1$, $0,$ and $-1$.
The case of generalized utility functions will be further explored in a future work.
Since our study focuses on the perspective of unlicensed users, we will use the terms ``user" and ``unlicensed user" interchangeably in the following analysis.

\subsection{Key game concepts}

\begin{definition}[\textbf{Social Welfare}]
The \textbf{social welfare} $\sum_{n=1}^N U_n({\boldsymbol x})$ of a strategy profile ${\boldsymbol x}$ is  the sum of all players' utilities.
\end{definition}

\begin{definition}[\textbf{Social Optimum}]
A strategy profile ${\boldsymbol x}$ is a social optimum when it maximizes social welfare.
\end{definition}

\begin{definition}[\textbf{Better Response Update}]
The event where a player $n$ changes its choice of strategy from $x_{n}$ to $c$ is a better response update if and only if $U_{n}(c,\boldsymbol{x}_{-n})>U_{n}(x_n,\boldsymbol{x}_{-n}),$ where \rev{we write the argument of the function as $\boldsymbol{x} = (x_n, \boldsymbol{x}_{-n})$ with} $\boldsymbol{x}_{-n}=(x_1,\edi{\ldots},x_{n-1},x_{n+1},\edi{\ldots},x_N)$ \rev{representing} the strategy profile of all players except player $n$.
\end{definition}

\begin{definition}[\textbf{Pure Nash Equilibrium}]\label{def:PNE}
A strategy profile ${\boldsymbol x}$ is a pure Nash equilibrium if no players at ${\boldsymbol x}$ can perform a better response update, i.e., $U_{n}(x_n,\boldsymbol{x}_{-n})\geq U_{n}(c,\boldsymbol{x}_{-n})$ for any $c\in\tilde{\mathcal{C}}$ and $n\in\mathcal{N}$.
\end{definition}

\begin{definition}[\textbf{Finite Improvement Property}]
A game has the finite improvement property if \emph{any} asynchronous better response
update process\footnote{Where no more than one player updates \rev{his} strategy at any given
time.} terminates at a pure Nash equilibrium within a finite number of updates.
\end{definition}

\subsection{Transformation to an equivalent interference threshold form}\label{physics2}

For \rev{the discussion convenience}, we will introduce an equivalent interference-threshold form of the QoS satisfaction game. The key idea is to relate a user's received congestion level with its QoS demand satisfaction.


Since the data rate function $Q_n^c(I^c)$ is non-increasing with the congestion level $I^c$, there must exist a critical threshold value $T_n^c$, such that $Q_n^c(I^c)\geq D_n$ if and only if the congestion level $I^c\leq T_n^c$. Formally, given a pair of $(Q_n^c,D_n)$, we shall define the threshold $T_n^c$ of channel $c$ with respect to user $n$ to be an integer such that
\begin{itemize}
\item \rev{If} $Q_n^c(I^c)<D_n$ for each $I^c \in \mathcal{N},$ then $T_n^c = 0$ \rev{(hence user $n$'s QoS demand can never be satisfied on channel $c$ even if it is the only user on this channel)},
\item \rev{If} $Q_n^c(I^c)>D_n$ for each $I^c \in \mathcal{N},$ then $T_n^c = N+1$ {(hence user $n$'s QoS demand is always satisfied on channel $c$ even if all users use this channel) \footnote{\edi{It is possible to set $T_n^c$ to be any number greater than $N$ while still  satisfy condition (\ref{eq:tf}). The reason of choosing $T_n^c = N+1$ is to bound the differences between thresholds, which helps the proof of fast convergence of the distributed algorithm in Theorem \ref{graphical}.}}},
\item \rev{Otherwise} $T_n^c$ is equal to the maximum integer $I^c \in \mathcal{N}$ such that $Q_n^c(I^c) \geq D_n$.
\end{itemize}
These conditions guarantee that
\begin{align}
Q_n^c(I^c) \geq D_n \Leftrightarrow I^c \leq T_n^c \label{eq:tf}.
\end{align}

We can then express a QoS satisfaction game $g=(\mathcal{N}, \mathcal{C}, (Q_n^c)_{n \in \mathcal{N}, c \in \mathcal{C}}, (D_n)_{n \in \mathcal{N}})$ in the interference threshold form $g'=(\mathcal{N}, \mathcal{C}, (T_n^c)_{n \in \mathcal{N}, c \in \mathcal{C}})$. And the utility of user $n$ can be computed accordingly as
\begin{equation}
  U_n(\boldsymbol{x})=\begin{cases}
    1, & \text{if $x_n \neq 0$ and $I^{x_n}({\boldsymbol x}) \leq T_n^{x_n}$},\\
    0, & \text{if $x_n = 0$},\\
    -1, & \text{if $x_n \neq 0$ and $I^{x_n}({\boldsymbol x}) > T_n^{x_n}$}.\\
  \end{cases} \label{eq:utility}
\end{equation}

The interference threshold transformation reduces the size of parameters by replacing $(Q_n^c,D_n)$ with $T_n^c$. Moreover, the result in (\ref{eq:tf}) ensures that the original game $g$ is equivalent to the game $g'$, since the utility $U_n(\boldsymbol x)$ received by player $n$ in $g$ is the same as that received by player $n$ in $g'$ for every strategy profile $\boldsymbol x$ and player $n$. For the rest of the paper, we will analyze the QoS satisfaction game in the interference threshold form. \edi{Note that Equations (\ref{uo}) and (\ref{eq:utility}) are equivalent. It is just that we write the latter expression in terms of thresholds.}

\section{Properties of The QoS Satisfaction Game}\label{results}

Now we explore the properties of QoS satisfaction games, including the existence of pure Nash equilibria and the finite improvement property.  We shall also describe the conditions under which a social optimum is also a pure Nash equilibrium. 

\subsection{Characterization of pure Nash equilibria}\label{nepic}

Each player is \rev{either} satisfied or dormant when a game is at pure Nash equilibrium. To see this, consider a strategy profile ${\boldsymbol x}$ where a player $n$ is suffering. Now the action where player $n$ changes its strategy to the dormant state $0$ is a better response update \rev{for this user}. Since suffering users can always do better response updates, \rev{such a strategy profile $\boldsymbol{x}$ cannot be a pure Nash equilibrium}.

Next we show in Theorem \ref{completeconvergence} that every QoS satisfaction game has the finite improvement property (\rev{which is a sufficient condition for} the existence of a pure Nash equilibrium).

\begin{theorem}\label{completeconvergence}
Every $N$-player QoS satisfaction game has the finite improvement property. Moreover, any asynchronous better response update process is guaranteed to reach a pure Nash equilibrium with no more than $4N + 3N^2$ asynchronous better response updates (irrespective of the initial strategy profile, or the order in which the players update).
\end{theorem}


Theorem \ref{completeconvergence} is a direct consequence of the more general result Theorem \ref{graphical} in  Section \ref{spatial}.
 Theorem \ref{completeconvergence} is very important, because it implies that the general QoS satisfaction games (with heterogenous channels and users) can self organize into a stable state effectively. This fact allows us to design the distributed QoS satisfaction algorithm in Section \ref{dis}, which has a fast convergent property. \edi{Although Theorem \ref{completeconvergence} shows that pure Nash equilibria can be found relatively easily, it does not offer any insight into how to select the most beneficial pure Nash equilibria. Equilibrium selection seems to be a difficult problem in the general case. However, we show how to find pure Nash equilibria which are social optimum for special cases in Subsections \ref{homouse} and \ref{homochan}.}

\subsection{Finding a social optimum is NP hard}\label{hard}

Although Theorem \ref{completeconvergence} implies that pure Nash equilibria are easy to construct, it turns out that finding a social optimum can be extremely challenging.

\begin{theorem}\label{nphard}
The problem of finding a social optimum of a QoS satisfaction game is NP hard.
\end{theorem}


\edi{The problem of finding a social optimum of a QoS satisfaction game has some resemblance to the Knapsack problem (where items have different weights and values, and the objective is to maximize the value of items chosen without exceeding a given total weight threshold). The key difference is that the thresholds in our problem are associated with the players/items we are choosing, and there are multiple channels/knapsacks to allocate our players to. Our proof (given in Appendix \ref{Proof1}) is based upon showing} that the 3-dimensional matching decision problem (which is well known to be NP complete \cite{3DM}) can be reduced to the problem of finding a social optimum of a QoS satisfaction game where thresholds $T_n^c \in \{1,3 \}$ for each $n$ and $c$. Theorem \ref{nphard} provides the major motivation for our game theoretic study, because it suggests that \rev{the} centralized \rev{spectrum sharing problem} is fundamentally difficult. It therefor makes sense to explore decentralized alternatives such as a game based \rev{spectrum sharing}.

\subsection{Price of anarchy}\label{poasec}

Although Theorem \ref{nphard} suggests that finding an optimal strategy profile can be very difficult, we do know from Theorem \ref{completeconvergence} that pure Nash equilibria can be found with relative ease. This naturally raises the question of how the social welfare of pure Nash equilibria compare to the maximum possible social welfare. In other words, \emph{how much social welfare can be lost by allowing the players to organize themselves, rather than being directed to a social optimum?}

To gain insight into this issue, we study the price of anarchy \cite{PoA}. Recall that $\tilde{\mathcal{C}}^N$ is the set of strategy profiles of our game. Let $\Xi \subseteq \tilde{\mathcal{C}}^N$ denote the set of pure Nash equilibria of our game. Note that Theorem \ref{completeconvergence} implies that $\Xi$ is non-empty. Now the \textbf{price of anarchy}
\begin{equation} \label{poadfn} \poa = \frac{\max\{\sum_{n =1}^N U_n({\boldsymbol x}) : {\boldsymbol x} \in \tilde{\mathcal{C}}^N \}}{\min\{\sum_{n =1}^N U_n({\boldsymbol x}) : {\boldsymbol x} \in \Xi \}}, \end{equation}
is defined to be the maximum social welfare of a strategy profile, divided by the minimum welfare of a pure Nash equilibrium. The social welfare of a system at a pure Nash equilibrium can be increased by at most $\poa$ times by switching to a centralized solution.

\begin{theorem}\label{poa}
\rev{Consider} a QoS satisfaction game $(N, C, (T_n^c)_{n \in \mathcal{N}, c \in \mathcal{C}})$, where $T_n^c \geq 1$ for each player $n$ and each channel $c$. \rev{The $\poa$ of} this game satisfies
 \begin{equation} \label{poaTH} \poa \leq \min \left\{ N , \frac{\max\{T_n^c: n \in \mathcal{N}, c \in \mathcal{C} \}}{\min\{T_n^c: n \in \mathcal{N}, c \in \mathcal{C} \}} \right\}. \end{equation}
\end{theorem}


The proof is given in Appendix \ref{Proof2}. \rev{The constraint $T_n^c \geq 1$ insures that some player will be satisfied in every pure Nash equilibrium of the game, and avoid the possibility of the $\poa$ involving ``division by zero''.} Theorem \ref{poa} implies that the performance of every pure Nash equilibrium will be close to optimal when the minimum threshold of a user-channel pair is close to the maximum threshold of a user-channel pair. This is a very significant result, when one considers that pure Nash equilibria can be easily reached by better response updates (Theorem \ref{completeconvergence}) while finding social optima is NP hard (Theorem \ref{nphard}). Motivated by Theorem \ref{poa}, we next study two special cases of QoS satisfaction games with homogenous settings, i.e., homogeneous users and homogenous channels. In both cases, the social optimum can be actually achieved at a pure Nash equilibrium.

\subsection{QoS satisfaction games with homogenous users}\label{homouse}

We first study the case of homogeneous users. We say that a QoS satisfaction game has \textbf{homogenous users} when $T_1^c = T_2^c = \ldots =T_N^c$, for each $c \in \mathcal{C}$ (i.e., each player has the same threshold for any channel $c$). This corresponds to the case that all users have the same data rate function $Q_n^c$ on the same channel $c$ (but they may have different data rates on different channels) and the same demand $D_n$. For example, spectrum sharing in a network of RFID tags in a warehouse may correspond to such a QoS satisfaction game, because every device experiences the same environment and requires a similar data rate to operate.

When discussing QoS satisfaction games with homogenous users, we drop the subscripts and use $T^c$ to denote the common threshold of all players on channel $c$.  Since users are homogenous,
we only need to keep track of how many users choose each channel in order to describe the game dynamics. Next we will show that any pure Nash equilibrium in a QoS satisfaction games with homogenous users is also a social optimum.


\begin{theorem}\label{homplayequiv}
Let ${\boldsymbol x}$ be a strategy profile of a QoS satisfaction game with $N$ homogenous users and $C$ channels, with thresholds $T^1,T^2,\ldots,T^C$. The following three statements are equivalent:
\begin{enumerate}
\item ${\boldsymbol x}$ is a pure Nash equilibrium;
\item There are no suffering users in ${\boldsymbol x}$ and the number of satisfied users is $\min\{N, \sum_{c=1}^C T^c\}.$
\item ${\boldsymbol x}$ is a social optimum.
\end{enumerate}
\end{theorem}

The proof is given in Appendix \ref{Proof5}. Theorems \ref{completeconvergence} and \ref{homplayequiv} together imply that \rev{any} sufficiently long asynchronous better response updating sequence will converge to a social optimal in polynomial time when the game has homogenous users. Moreover, Theorem \ref{homplayequiv} implies that when $\sum_{c=1}^C T^c\geq N$, there exists a satisfaction equilibrium \cite{satform} where all the players can be satisfied.

\subsection{QoS satisfaction games with homogenous channels}\label{homochan}

We next consider the case that the \emph{channels} are homogenous. We say a QoS satisfaction game has \textbf{homogenous channels} when $T_n^1 = T_n^2 = \ldots =T_n^C$, for each user $n$ (i.e., all channels have the same threshold from any player's perspective). This corresponds to the case that each user $n$ has the same data rate function $Q_n^c$ on all the channels, but different users may have different demands $D_n$. QoS satisfaction games with homogenous channels are highly relevant, because technologies such as frequency interleaving can been adopted in many wireless systems such as IEEE 802.11g networks \cite{IEEE802} to make channels homogeneous (i.e., having the same bandwidth and experiencing frequency flat fading).

When discussing QoS satisfaction games with homogenous channels, we drop the superscripts and use $T_n$ to denote the common threshold of  player $n$ for all channels.  The update process can reach a pure Nash equilibrium according to Theorem \ref{completeconvergence}.

We next discuss the optimality of pure Nash equilibria. Firstly note that, a pure Nash equilibrium may not be a social optimal. For example, let us consider a game of six users, with thresholds $T_1 = T_2 = 2$, $T_3 = T_4 = T_5 = T_6 = 4$, and two channels. The game has a pure Nash equilibrium ${\boldsymbol x} = (0,0,1,1,2,2)$ with four satisfied users, which is not a social optimum. \rev{The strategy profile ${\boldsymbol y} = (1,1,2,2,2,2)$, where all six users are satisfied, is a social optimum.}

Second, a social optimum may not be a pure Nash equilibrium. We take the game with six users and thresholds $T_1 = T_2 = 2$, $T_3 = T_4 = T_5 =3$, $T_6 = 4$, and two channels as an example. The game has a social optimum ${\boldsymbol x} = (1,1,2,2,2,0)$ (with five satisfied users), which is not a pure Nash equilibrium because user $6$ can do a better response update by switching to channel $1$.

Surprisingly, there always exists a pure Nash equilibrium that is a social optimum for a game with homogenous channels. Moreover, we present an algorithm (Algorithm \ref{d}) that always generates a social optimum which is a pure Nash equilibrium. The key idea of the algorithm is to prioritize channel allocation according to users' thresholds (i.e., the more severe congestion a user can tolerate, the higher priority it will get in channel allocation).

\begin{figure}[tt]
\centering
\includegraphics[scale=0.5]{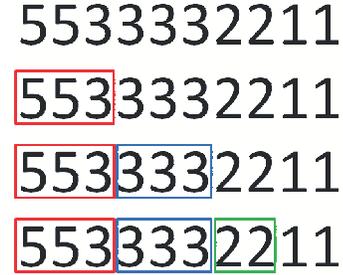}
\caption{\rev{An illustration of Algorithm \ref{d} in action. The ten users have thresholds $(T_1,T_2,T_3,T_4,T_5,T_6,T_7,T_8,T_9,T_{10}) = (5,5,3,3,3,3,2,2,1,1)$. Each row represents a strategy profile. Each number represents a user with that number equal to its threshold. The top (first) row represents the initial strategy profile $\boldsymbol{x^0}$ of our algorithm (where all players are dormant). The second, third and fourth rows represent the strategy profiles $\boldsymbol{x^3}$, $\boldsymbol{x^6}$ and $\boldsymbol{x^8}$. The boxes within a given row represent channel allocations, in the sense that the users contained within the leftmost (red) box are using channel $1$, the users contained within the central (blue) box are using channel $2$ and the users contained within the rightmost (green) box are using channel $3$. The bottom row represents strategy profile ${\boldsymbol x^8} = {\boldsymbol x^{10}= (1,1,1,2,2,2,3,3,0,0)}$, which is the output of the algorithm (within which $8$ players are satisfied).}}
\label{MergedR}
\end{figure}

\rev{Algorithm \ref{d} is a centralized algorithm that demonstrates the existence of a pure Nash equilibrium which is a social optimum. The distributed algorithm that globally converges to a Nash equilibrium (not necessarily socially optimal) will be discussed in Subsection \ref{dis}.} Algorithm \ref{d} begins by making all players dormant. The players are then updated
\rev{one by one} in \rev{the} order of descending \rev{thresholds}. When a player is \rev{updated, it changes} to the lowest indexed channel which will satisfy \rev{it}. \rev{If there are no channels that can satisfy this player, then the algorithm will not further change players' channel choices,  since all higher indexed players will not be able to find channels to satisfy them as they have even lower interference thresholds}.\com{please double check my change in the previous sentence.}). Figure \rev{\ref{MergedR}} illustrates a particular example of Algorithm \ref{d} running. We show in Theorem \ref{algodworks} that Algorithm \ref{d} is guaranteed to generate a pure Nash equilibrium that is also a social optimum.

\comg{Thats fine}


\begin{algorithm}[Http]
\label{homochanalgo}
\caption{Finds \rev{a pure Nash equilibrium that is a social optimum for} a game with homogenous channels.}
\label{d}
\KwIn{A QoS satisfaction game with $C$ \rev{homogenous} channels and $N$ players, \rev{who} have thresholds $T_1 \geq T_2 \geq \edi{\ldots}. \geq T_N$.}
 \KwOut{A social optimum which is a pure Nash equilibrium.}
Let ${\boldsymbol x^0} = (x^0_1,x^0_2,\edi{\ldots},x^0_N) = (0,0,\edi{\ldots},0)$\\
{\For{$n=1$ \KwTo $N$}
{
\If{$\exists c \in \mathcal{C} :I^c(\boldsymbol{x^{n-1}})<T_n$}{
Let $c^* = \min\{ c \in \mathcal{C}: I^c(\boldsymbol{x^{n-1}})<T_n \}$\\
Let ${\boldsymbol x^n}= (x^{n-1}_1,\edi{\ldots},x^{n-1}_{n-1},c^*,x^{n-1}_{n+1},\edi{\ldots},x_N^{n-1})$
}\Else{Let ${\boldsymbol x^n} = {\boldsymbol x^{n-1}}$}
}}
\Return ${\boldsymbol x^N}$
\end{algorithm}


\begin{theorem}\label{algodworks}
Algorithm \ref{d} has a complexity of $O(C N^2)$ and generates a strategy profile that is both a social optimum and a pure Nash equilibrium of a QoS satisfaction game with $C$ homogeneous channels and $N$ users.
\end{theorem}



We provide the proof of Theorem \ref{algodworks} in Appendix \ref{ProofSketch}. Next Theorem \ref{xuopt} gives a sufficient condition for the existence of \rev{a} strategy profile where all players are satisfied, in a QoS satisfaction game with homogeneous channels (please refer to Appendix \ref{Proof3} for the proof).

\begin{theorem}\label{xuopt} If $T_n \geq \lceil\frac{N}{C}\rceil$ holds for every user $n$ in the QoS satisfaction game with $C$ homogeneous channels and $N$ users, then there is a strategy profile ${\boldsymbol x}$ within which every user is satisfied \rev{(which is a pure Nash equilibrium)}.
\end{theorem}




\section{Spatial QoS satisfaction Game}\label{spatial}

In all the games considered so far, we have \rev{assumed} that every pair of users are close enough to cause congestion to each other, when they use the same channel. However, in reality only nearby users of the same channel will cause congestion to one another, and distantly spaced users may access the same channel without degrading each \rev{other's} QoS. This \rev{is} known as \emph{spatial reuse} -- where the same piece of spectrum can be used by many distantly separated users without detrimental effects.

The protocol interference model \cite{localbroad} is a commonly used \rev{model} to approximate how the \rev{positions} of users \rev{affect} their communication performance. The idea behind the protocol interference model is to construct \rev{an} \emph{interference graph,} \rev{where} vertices represent players \rev{(wireless users), and an undirected edge connecting two players represents that these two players} are within interference range of one another (\rev{hence} they can generate interference to each other \rev{if transmitting on the same channel}). By using an interference graph $G$ to represent which vertices are close enough to interfere with each other, one may view the spectrum sharing problem as a game on a graph.

In this game, one may determine whether the QoS demand of a user is satisfied by counting the number of neighbors \rev{it has}, which are using the same channel as \rev{itself}. This corresponds to a generalization of the QoS satisfaction game where we account for the spatial positioning of the users.


Let us define a \textbf{spatial QoS satisfaction game} to be a quadruple $(\mathcal{N}, \mathcal{C}, (T_n^c)_{n \in \mathcal{N}, c \in \mathcal{C}}, G)$ where:
\begin{itemize}

\item $\mathcal{N}$, $\mathcal{C}$, and $T_n^c$ are the set of players/users, channels, and thresholds, respectively, which are the same as those introduced in Section \ref{physics2}.

\item $G= (\mathcal{N}, \rev{\mathcal{E}})$ is an undirected and unweighted graph, with a vertex set equal to the set of players $\mathcal{N}$, and an edge set $\rev{\mathcal{E}}$. We refer to $G$ as the \textbf{interference graph}. The interpretation of $G$ is that there is an edge $\{n,m\} \in \rev{\mathcal{E}}$ if and only if users $n$ and $m$ are close enough to cause congestion to each other \rev{when transmitting on the same channel}. We can apply the interference estimation methods in \cite{niculescu2007interference,Xia2013} to obtain the interference graph.
\end{itemize}

As before, a strategy profile ${\boldsymbol x} = (x_1,x_2,\edi{\ldots},x_N)$ is where each player $n$ chooses a strategy $x_n \in \tilde{\mathcal{C}}$. Let us define the \textbf{neighborhood} of player $n$, to be $\Ne (n) = \{m : \{n,m\} \in \rev{\mathcal{E}}\} \cup \{n\}$. In other words $\Ne(n)$ is the set of all players which are linked to, or identical to $n$. We let the neighborhood of a player contain the player itself \rev{just} for \rev{the} notational convenience.

\begin{figure}[tt]
\centering
\includegraphics[scale=0.45]{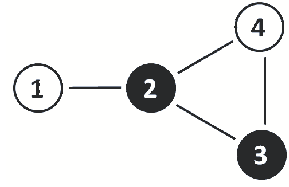}
\caption{A strategy profile in a spatial QoS satisfaction game. Each player (vertex) has chosen some channel (color). Player $2$ is linked to one other user of the black channel $b$, so that the congestion level $I_b^2({\boldsymbol x})$ of the black channel for player $2$ is two. If $T_b^2 \geq 2$ then player $2$ will be satisfied in this strategy profile.}
\label{satgraph}
\end{figure}

Let use define the \textbf{local congestion level} of channel $c$ for player $n$ in strategy profile ${\boldsymbol x}$ to be
$$I_n^c({\boldsymbol x}) = |\{m \in \Ne(n) : x_m = c \}|.$$ In other words, $I_n^c({\boldsymbol x})$ denotes the number of players within a graph-distance $1$ of $n$ that are using the same channel as $n$. The utility player $n$ gets in strategy profile ${\boldsymbol x}$ is defined in a similar way to Equation (\ref{eq:utility}), from Subsection \ref{physics2}. We illustrate a spatial QoS satisfaction game in Figure \ref{satgraph}.



\begin{theorem}\label{graphical}
Every $N$-players spatial QoS satisfaction game has the finite improvement property. Moreover, any asynchronous better response update process will reach a pure Nash equilibrium within $4N + 3N^2$ asynchronous better response updates (irrespective of the initial strategy profile, or the order in which the players update).
\end{theorem}


The proof is given in Appendix \ref{Proof4}. Theorem \ref{graphical} \rev{is the most powerful result in this paper}, for it implies that every spatial QoS satisfaction game, with heterogenous players and heterogenous channels has the finite improvement property. The type of QoS satisfaction games we defined in Section \ref{model} can be considered \rev{as special cases} of spatial QoS satisfaction games within which the interference graph is a complete graph. For this reason Theorem \ref{completeconvergence} can be considered to be \rev{a corollary} of Theorem \ref{graphical}. Theorem \ref{graphical} shows that spatial QoS satisfaction games are a remarkable class of congestion games on graphs, because they may have heterogenous channels and users, and yet they always have the finite improvement property. If one considers the slightly more general class of congestion games on graphs \cite{cem} with arbitrary non-increasing utility functions, then one can easily find example games which do not even have pure Nash equilibria -never mind the finite improvement property. For example, a congestion game on a graph with $5$ players and $3$ resources, without any pure Nash equilibria is exhibited in \cite{cem}.

\section{Distributed algorithm and simulations}\label{dist}

\subsection{Distributed QoS satisfaction algorithm}\label{dis}



In this section we propose a distributed QoS satisfaction algorithm for achieving pure
Nash equilibria of general (spatial) QoS satisfaction games. The key
idea is to utilize the finite improvement property and let one user improve its channel selection at a time. In order to describe the QoS satisfaction game purely in terms of channel selection, we may regard the dominant state $0$ as an addition \emph{virtual} channel, which always gives users a utility of $0$.


We consider a time-slotted system.
Each time slot $t$ consists of the following two parts:
 \ \\
1) \textbf{Spectrum Access}: each user $n$ contends to access the chosen channel
$x_{n}$ according to some medium access control (MAC) mechanism.
For the initialization, we assume that all users are dormant, and use strategy $0$.
\ \\
2) \textbf{Channel Update Contention}: We exploit the finite improvement property by having
one user carry out a channel update at each
time slot. In this part, we let users who can improve their channel selections compete
for the channel update opportunity in a distributed manner. More specifically, each user $n$ first computes its set of \rev{best responses (which is the set of strategies which maximize (and increase) $n$'s utility).}
\begin{align*}
\mathcal{B}_{n}(\boldsymbol x)= \{ & c^*:  c^*= \arg\max_{c \in \tilde{\mathcal{C}}} U_n(c,\boldsymbol{x}_{-n}) \mbox{ and} \\
& U_n(c^*,\boldsymbol{x}_{-n})>U_n(\boldsymbol{x})\}.
\end{align*}
If $\mathcal{B}_{n}(\boldsymbol x)\neq\varnothing$ (i.e., user $n$ can improve), then user $n$ will contend for the channel update opportunity. Otherwise, user $n$ will not contend and will adhere to the original channel selection $x_{n}$ at next time slot.


For the channel update contention, for example, we can adopt the backoff-based mechanism
by setting the time length of channel update contention as $\tau^*$.
Each contending user $n$ first generates a backoff time value $\tau_{n}$ according
to the uniform distribution over $[0,\tau^*]$ and waits until the
backoff timer expires. When the timer expires, if the user has not
received any updating messages from other users yet, the user will
randomly select a channel $c^{*}\in \mathcal{B}_{n}(\boldsymbol x)$ and broadcast
an updating message over the common control channel to indicate that
it will update its channel selection to \rev{$c^{*}$ at the beginning of the} next time slot.

According to the finite improvement property in Theorem \ref{graphical}, the algorithm will converge
to a pure Nash equilibrium of a general spatial QoS satisfaction game in polynomial time.


\begin{algorithm}[Http]
\caption{Distributed QoS satisfaction algorithm}
\label{dd}

\textbf{initialization:} each user $n$ chooses channel $x_{n}=0$.\\
\For{each user $n$ and each time slot $t$}{
access the chosen channel $x_{n}$.\\
compute the set of best response channel selections $\mathcal{B}_{n}(\boldsymbol x)$.\\
\eIf{$\mathcal{B}_{n}(\boldsymbol x)\neq\varnothing$}{
contend for the channel update opportunity.\\
\eIf{win the channel update contention}{
choose a channel $c^{*}\in \mathcal{B}_{n}(\boldsymbol x)$ randomly for next time slot.\\
broadcast the updated channel selection $c^{*}$ to other users.\\
}{
choose the original channel $x_{n}$ for next time slot.\\
}
}{
choose the original channel $x_{n}$ for next time slot.\\
}
update the channel selections $\boldsymbol{x}_{-n}$ of other users once an updating message is received.}
\end{algorithm}

\begin{figure*}[tt]
\begin{minipage}[t]{0.32\linewidth}
\centering
\includegraphics[scale=0.35]{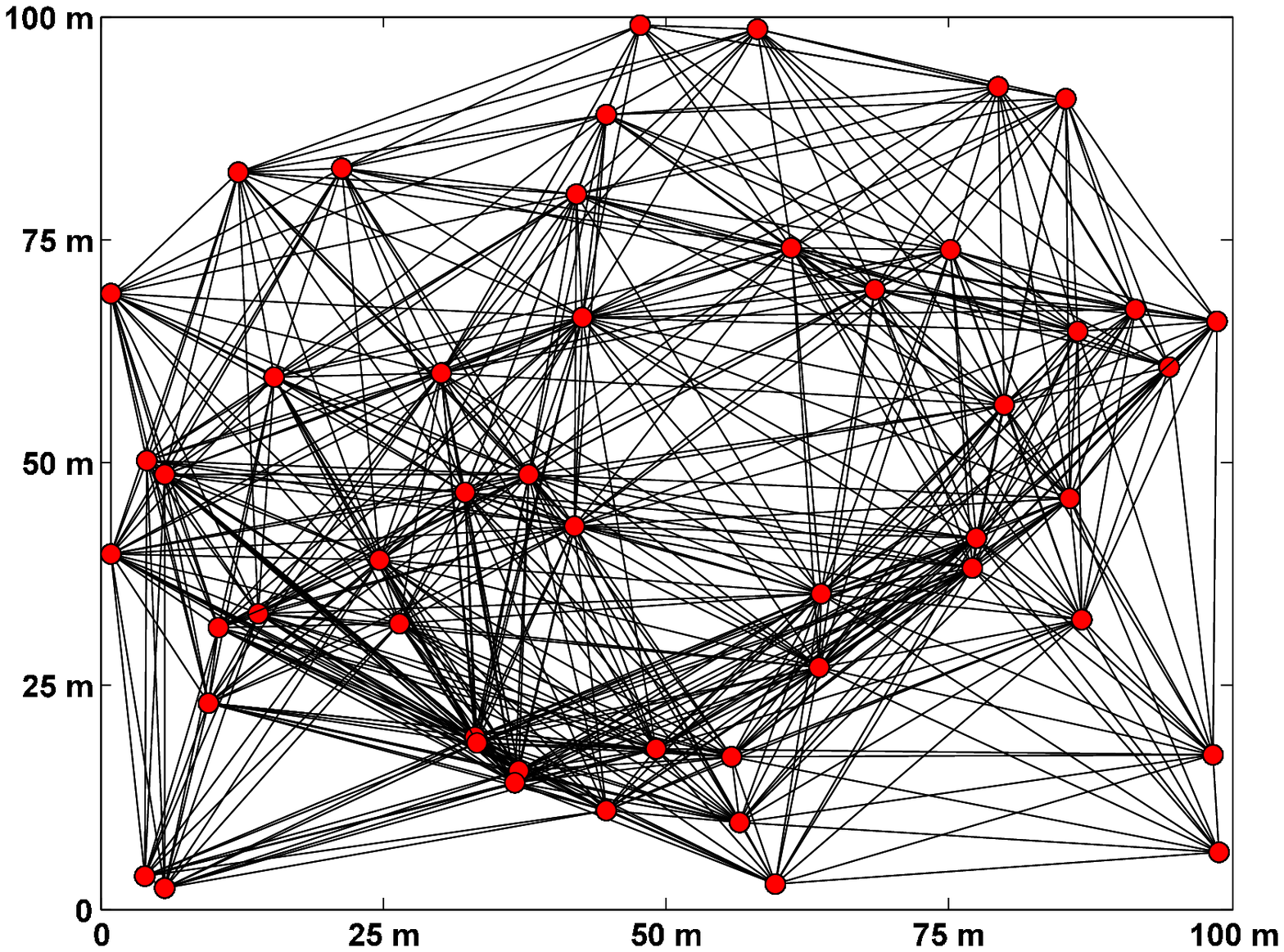}
\caption{\label{simul0}Interference graph generated by $N=50$ random scattered users over a $100 \m \times 100 \m$ region.  Each user (represented by a dot) has an interference range of $50 \m$. Two users are linked by an edge if and only if they are within each other's interference range.}
\end{minipage}
\hfill
\begin{minipage}[t]{0.33\linewidth}
\centering
\includegraphics[scale=0.45]{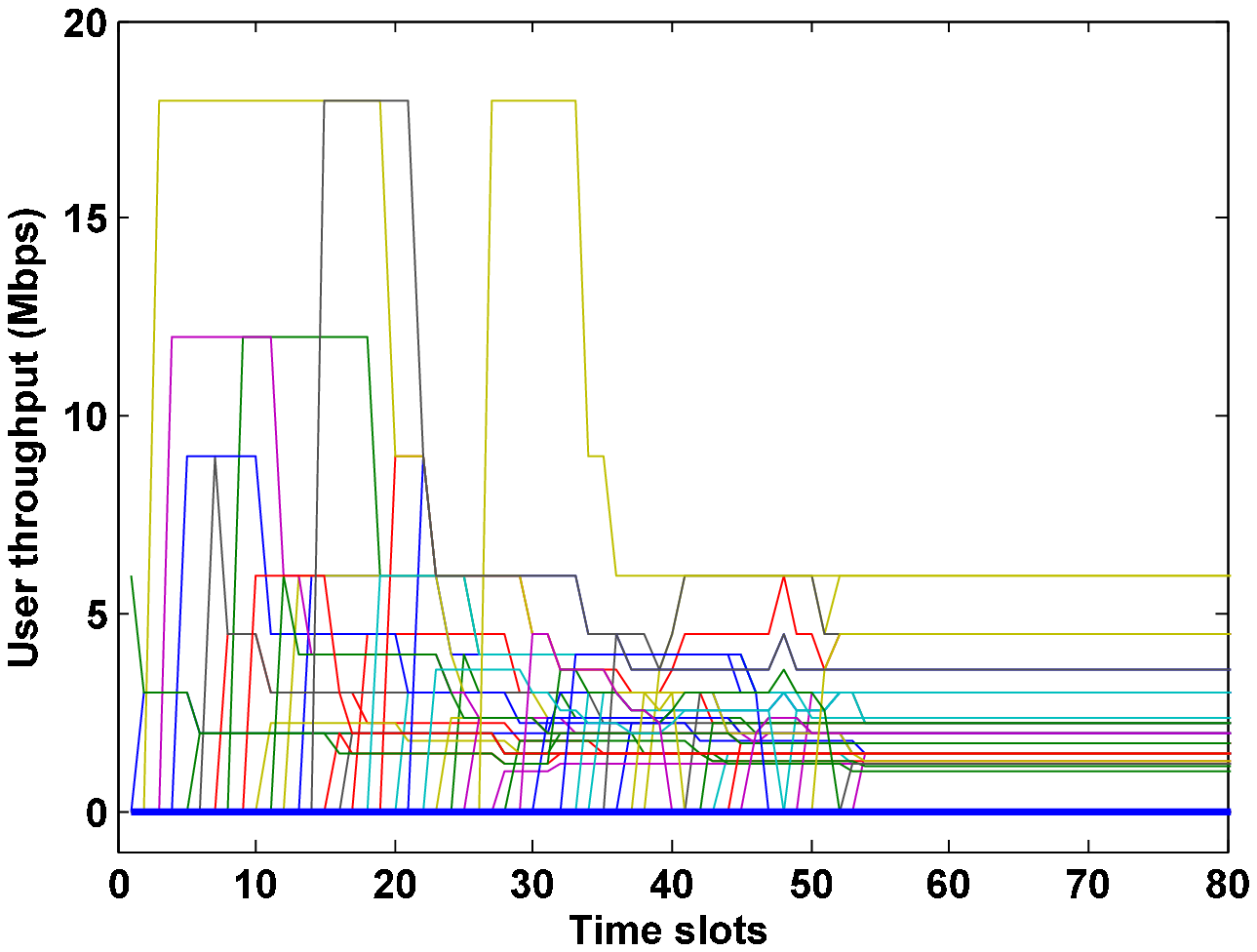}
\caption{\label{UserThr}Dynamics of users' throughputs by distributed QoS satisfaction algorithm. When the throughput of a user is zero, then the user is in the dormant state.}
\end{minipage}
\hfill
\begin{minipage}[t]{0.32\linewidth}
\centering
\includegraphics[scale=0.45]{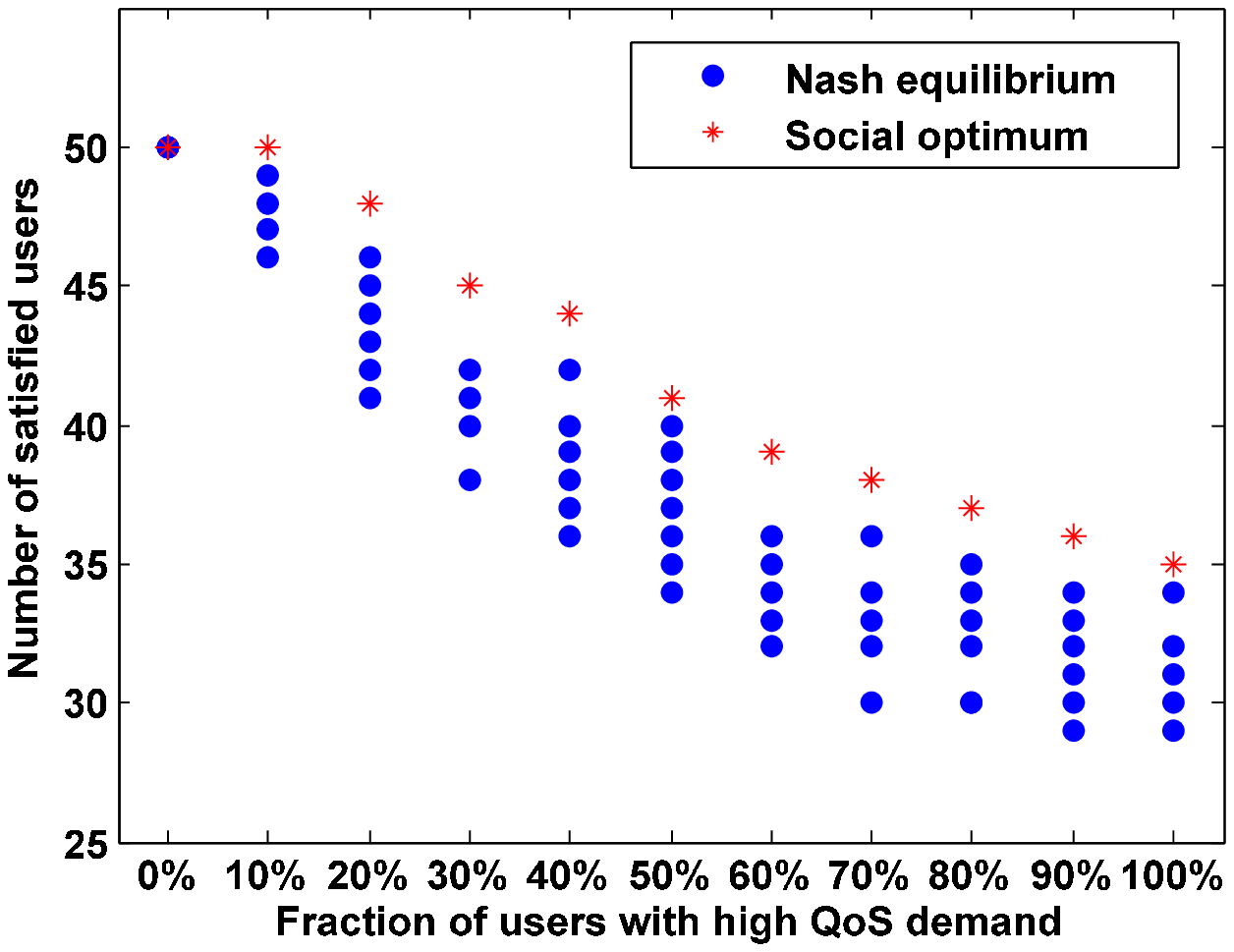}
\caption{\label{simul1}Number of satisfied users at pure Nash equilibria and social optima when the number of users is $N=50$, and the fraction of users with a high QoS demand ranges from $0\%$ to $100\%$, respectively. }
\end{minipage}
\end{figure*}

\begin{figure*}[tt]
\begin{minipage}[t]{0.48\linewidth}
\centering
\includegraphics[scale=0.55]{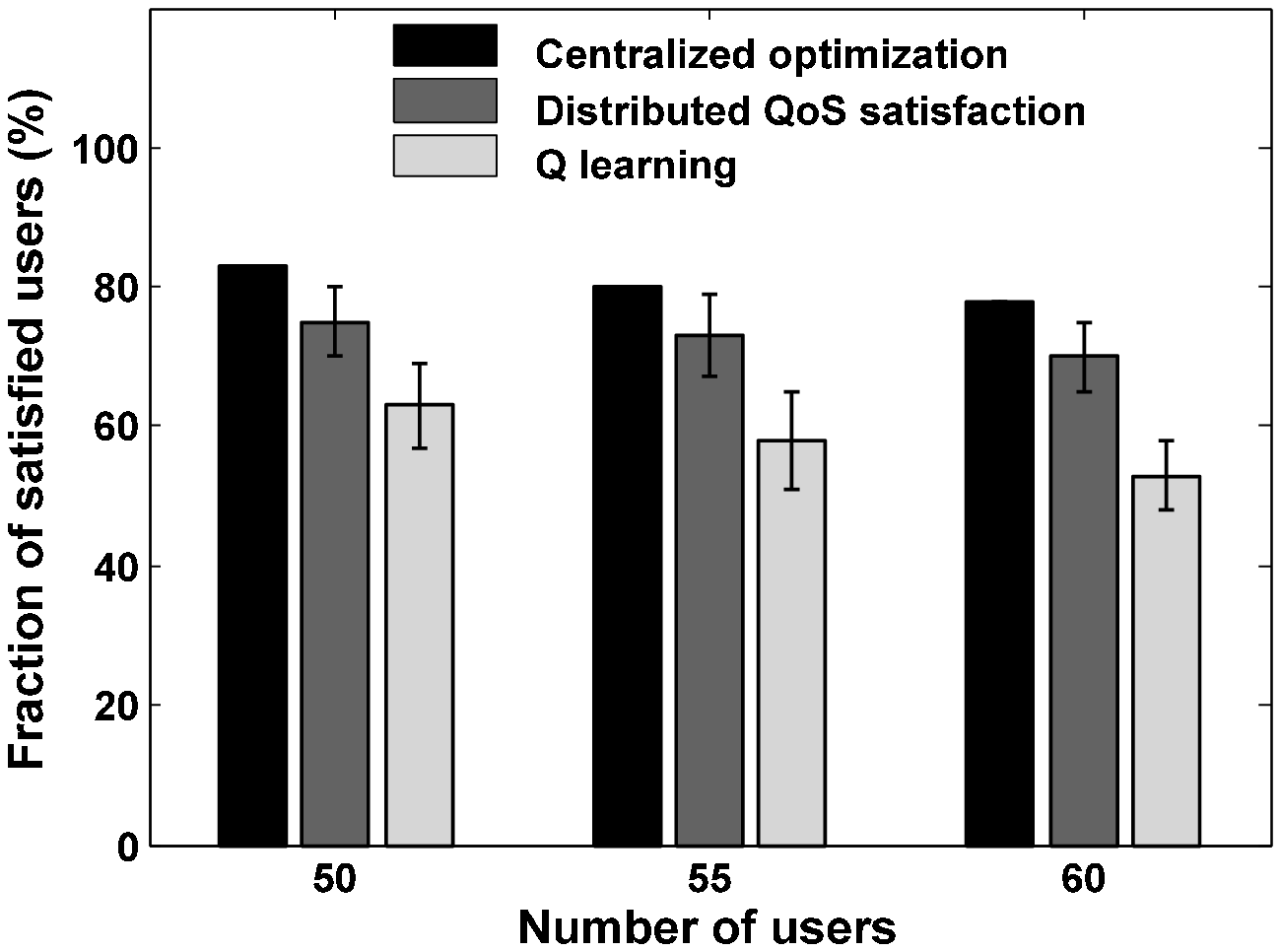}
\caption{\label{Comparison}Performance comparison of the centralized optimization, distributed QoS satisfaction algorithm, and Q-learning mechanism. }
\end{minipage}
\hfill
\begin{minipage}[t]{0.48\linewidth}
\centering
\includegraphics[scale=0.53]{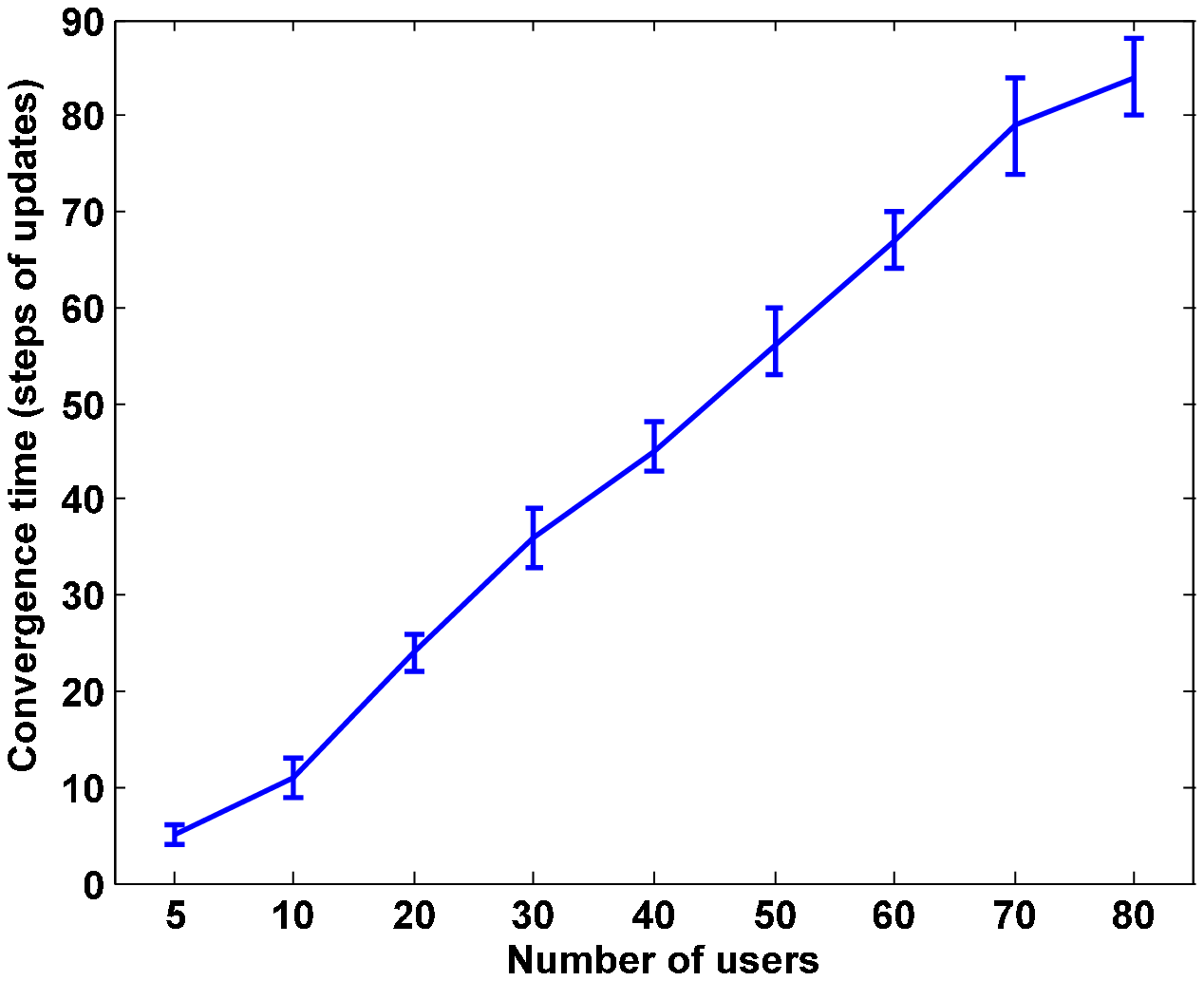}
\caption{\label{simul2}The convergence time of Algorithm \ref{dd} with  $N=5,10,20,\edi{\ldots},80$ users, and half of the users have a high QoS demand.}
\end{minipage}
\end{figure*}

\subsection{Numerical Results}\label{sim}

We now evaluate the proposed distributed QoS satisfaction algorithm by simulations. We consider a spectrum sharing network of $C=4$ vacant channels, with the mean data rates $B^c_n$ of $6, 9, 12, 18$ Mbps, respectively, which are standard operating data rates in IEEE 802.11g systems \cite{IEEE802}. Multiple users are randomly scattered over a $100 \m \times 100 \m$ region (see Figure \ref{simul0} for an illustration). In the interference graph, a pair of users are linked by an edge when they are within $50 \m$ (the interference range) of each other (i.e., when they can generate interference to each other). We adopt the TDMA mechanism for the medium access control (MAC) and the data rate of user $n$ choosing a channel $c$ is given as $Q_n^c(I_n^c({\boldsymbol x}))=\frac{B^c}{I_n^c({\boldsymbol x})}$, where $I_n^c({\boldsymbol x})$ is the number of users of channel $c$ that are linked to $n$ upon the interference graph. We consider the scenario where users are running two different multimedia applications \rev{corresponding to} two types of QoS demands: low demand type $D_n=0.125 \mbps$ (i.e., listening to an online MP3 \rev{song} \cite{adobe}) and high demand type $D_n=3.5 \mbps$ (i.e., watching an online video \rev{with a resolution of 1080p} \cite{adobe}).

We first implement a simulation with $N=50$ users, and let the fraction of users with a high QoS demand vary from $0\%$ to $100\%$. We implement the distributed QoS satisfaction game solution in Algorithm \ref{dd}. Figure \ref{UserThr} shows the dynamics of users' throughputs, which demonstrates that the proposed distributed QoS satisfaction algorithm can converge to a pure Nash equilibrium. As a benchmark, we also compute the social optimum by the centralized optimization using Cross Entropy method, which is an advanced randomized searching technique and has been shown to be efficient in solving complex combinatorial optimization problems \cite{rubinstein2004cross}. The results are shown in Figure \ref{simul1}. The x-axis is the fraction of users having a high QoS demand, and y-axis describes how many users are satisfied at the solutions of pure Nash equilibria and social optima. Note that a QoS satisfaction game may have multiple pure Nash equilibria, and Algorithm \ref{dd} will randomly select one pure Nash equilibrium (since a random user will be chosen for channel selection update). We run the algorithm $20$ times for each game instance and plot the number of satisfied users at the obtained pure Nash equilibria. Figure \ref{simul1} shows  that both the performances of social optima and (the best and the worst) pure Nash equilibria decrease as the fraction of users of a high QoS demand increases. This is because that given the constant spectrum resources less users can be satisfied when more users have higher demands. Compared with the social optima, the performance loss by the best pure Nash equilibria and the worst pure Nash equilibria by Algorithm \ref{dd} are at most $7\%$ and $20\%$, respectively \rev{(not shown in the figure)}.  This demonstrates the efficiency of the pure Nash equilibria of QoS satisfaction games.

We implement another simulation with the number of users $N=50,55,$ and $60$ with half of the users having a high QoS demand. Upon comparison, we also implement the social optimum solution by centralized optimization and the decentralized spectrum access solution by Q-learning mechanism proposed in \cite{li2010multi}. We observe that the distributed QoS satisfaction algorithm can achieve up-to $32\%$ performance gain over the Q learning mechanism. Compared with the centralized optimization, the performance loss of the distributed QoS satisfaction algorithm is at most $10\%$. This demonstrates the efficiency of the proposed distributed QoS satisfaction algorithm. We next evaluate the convergence time of the distributed QoS satisfaction algorithm. Figure \ref{simul2} shows that the average convergence time increases linearly with the number of users $N$. This shows that the distributed QoS satisfaction algorithm scales well with the network size. This is critical since computing the social optimum of general QoS satisfaction games is NP-hard.

%
%
%
%

\section{Conclusion}\label{conc}

In this paper, we proposed a framework of QoS satisfaction games to model the distributed QoS satisfaction problem among wireless users. The game based solution is motivated by the observation that the centralized optimization problem of maximizing the number of satisfied users is NP hard. We have explored many aspects of QoS satisfaction games including the pure Nash equilibria and the price of anarchy. Our results reveal that selfish spectrum sharing can be a very effective way to allow users to meet their QoS demands. In particular, we have shown that our systems can always reach a pure Nash equilibrium in polynomial time, simply by having the users perform better response updates.

There are many other issues we wish to explore in the future. In particular, we wish to extend many of our results (such as those regarding the price of anarchy) to spatial QoS satisfaction games.  We also wish to explore the generalized QoS satisfaction games where different players receive different utilities for being satisfied.

\begin{appendix}


\subsection{Proof of Theorem \ref{nphard}}\label{Proof1}
In the following, we call the problem of finding a social optimum of the QoS satisfaction game as the QoS satisfaction problem for short. Before discussing the computational complexity of the QoS satisfaction
problem, we first introduce the definition of 3-dimensional
matchings.
\begin{definition}\label{3Dmatching}
Let $\mathcal{X},\mathcal{Y},$ and $\mathcal{Z}$ be three finite disjoint sets, and let $\mathcal{T}$ be a subset
of $\mathcal{X}\times \mathcal{Y}\times \mathcal{Z}$. That is, $\mathcal{T}\subseteq\{(x,y,z):x\in \mathcal{X}, y\in \mathcal{Y}, z\in \mathcal{Z}\}$. Now $\mathcal{M}\subseteq \mathcal{T}$ is
a 3-dimensional matching if the following holds: for any two distinct
triples $(x_{1},y_{1},z_{1})\in \mathcal{M}$ and $(x_{2},y_{2},z_{2})\in \mathcal{M}$, we have $x_{1}\neq x_{2}$, $y_{1}\neq y_{2},$
and $z_{1}\neq z_{2}$.
\end{definition}

We shall refer to an element $(x,y,z)\in \mathcal{T}$ as an \emph{edge}. The 3-dimensional matching decision problem is as follows. Suppose
that the set sizes satisfy $|\mathcal{X}|=|\mathcal{Y}|=|\mathcal{Z}|=I$. Given an input $\mathcal{T}$ with $|\mathcal{T}|\geq I$, decide whether there exists
a 3-dimensional matching $\mathcal{M}\subseteq \mathcal{T}$ with the maximum size $|\mathcal{M}|=I$. The
3-dimensional matching decision problem is a well-known NP-complete
problem \cite{3DM}. We then prove that the QoS satisfaction problem is NP-hard, by showing that given an oracle for solving the QoS satisfaction problem, the 3-dimensional matching decision problem can be solved in polynomial time.

From an instance of 3-dimensional matching $((\mathcal{X},\mathcal{Y},\mathcal{Z}),\mathcal{T})$ with $|\mathcal{X}|=|\mathcal{Y}|=|\mathcal{Z}|=I$
and $|\mathcal{T}|=J\geq I$, we can create an instance of QoS satisfaction
problem as follows. The set of channels is $\mathcal{T}$ (i.e.,
each edge $(x,y,z)\in \mathcal{T}$ is a channel) with the total number of channels is $|\mathcal{T}|=J$. Let set $\psi=\mathcal{X}\cup \mathcal{Y}\cup \mathcal{Z}$. We regard each element
$n\in\psi$ as a user $n$. We also introduce a new user
set $\phi$ that consists of $J-I$ additional users. The total number of users in both $\psi$ and $\phi$ is $3I+J-I=2I+J$. \com{these users are in different user set, and thus are different by definition.} Then we define
the threshold value $T_{n}^{m}$ as follows. For a user $n$ in
set $\psi$ on a channel $m=(x,y,z)$, we set $T_{n}^{m}=3$ if $n$ is an element of an edge $m$
in $\mathcal{T}$ (i.e., one of the following cases is true: $n=x$, or $n=y$, or $n=z$), and we set $T_{n}^{m}=1$ otherwise. For a user
$n$ in set $\phi$ on a channel $m=(x,y,z)$, we set $T_{n}^{m}=1$. Clearly, $3$ users can stay in a channel and satisfy their QoS demands simultaneously if and only if they forms an edge in
$T$.\com{Say set $\mathcal{T}$ includes edges like $(1,2,4)$ and $(1,3,5)$. But users 1, 2, 5 cannot appear in the same edge, since user $1$ has threshold of $3$ in channels $(1,2,4)$ and $(1,3,5)$ only, user 2 has threshold of $3$ in channel $(1,2,4)$, and user $5$ has threshold of $3$ in channel $(1,3,5)$. For other channels, users 1, 2, 5 have threshold of $1$.} Since each user can only select one channel, according to Definition \ref{3Dmatching}, given a channel allocation solution, the set of channels, each of which has $3$ satisfied users, hence correspond to a 3-dimensional matching in $\mathcal{T}$.  In this case, the QoS satisfaction problem has the optimal solution that
all the users are satisfied (i.e., the number of satisfied users on $J$ channels is $3I+J-I=2I+J$ including $I$
channels with each channel having $3$ satisfied users and $J-I$
remaining channels with each channel having $1$ satisfied user),\com{ there are J channels in total.} if and only if there exists a 3-dimensional matching $\mathcal{M}\subseteq \mathcal{T}$ that has the maximum size $|\mathcal{M}|=I$.\com{If the matching is not of the maximum size $I$, not all the users can be satisfied. }

Therefore, if we have an oracle to find the optimal solution
for QoS satisfaction problem, we can then check whether the number of satisfied users is $2I+J$. In this case, we can decide in a polynomial time $\mathcal{O}(1)$ whether there exists a 3-dimensional matching $\mathcal{M}\subseteq \mathcal{T}$ such that $|\mathcal{M}|=I$. That is, 3-dimensional matching decision problem is polynomially
reducible to the QoS satisfaction problem, and hence the QoS satisfaction problem is NP-hard.\hfill $\Box$


\subsection{Proof of Theorem \ref{poa}}\label{Proof2}


Before \rev{proving the} main result about price of anarchy, let us establish a useful lemma. Let $B(\boldsymbol{x})$ denote the number of satisfied users in a strategy profile $\boldsymbol{x}$.

\begin{lemma}\label{usefullemma}
Suppose that ${\boldsymbol x}^*$ is a social optimum, and ${\boldsymbol y}^*$ is a pure Nash equilibrium of a QoS satisfaction game. \rev{The} following statements are true:
\begin{enumerate}
\item There are no suffering users in ${\boldsymbol x}^*$ (i.e., ${\boldsymbol x}^*$ is natural).
\item We have
$\sum_{n =1}^N U_n({\boldsymbol {\boldsymbol x}^*}) = B({\boldsymbol x}^*) = \sum_{c=1} ^C I^c ({\boldsymbol x}^*).$
\item There are no suffering users in ${\boldsymbol y}^*$ (i.e., ${\boldsymbol y}^*$ is natural).
\item We have
$\sum_{n =1}^N U_n({\boldsymbol {\boldsymbol y}^*}) = \rev{ B({\boldsymbol y}^*)} = \sum_{c=1} ^C I^c ({\boldsymbol y}^*).$
\end{enumerate}
\end{lemma}

\textit{Proof of Lemma \ref{usefullemma}:}
Statement \rev{1)} holds because the social welfare of any strategy profile with a suffering user can be increased by making the suffering user dormant.

Statement \rev{1)} implies that \rev{for any $n$} we have $U_n({\boldsymbol x}^*) \in \{0,1\}$. Also we have $U_n({\boldsymbol x}^*)=1$ if and only if \rev{user} $n$ is satisfied in ${\boldsymbol x}^*$. It follows that $\sum_{n =1}^N U_n({\boldsymbol {\boldsymbol x}^*})$ \rev{equals} $B({\boldsymbol x}^*)$ \rev{which is} the number of satisfied players in ${\boldsymbol x}^*$. Moreover, since every non-dormant user is satisfied under ${\boldsymbol x}^*$, and $\sum_{c=1} ^C I^c ({\boldsymbol x}^*)$ equals the number of non-dormant users under ${\boldsymbol x}^*$, we must have $\sum_{c=1} ^C I^c ({\boldsymbol x}^*) = B({\boldsymbol x}^*)$. \rev{This finishes the proof of Statement \rev{2)}}.

To see that Statement \rev{3)} holds, note that any suffering user can do a better response update by becoming dormant. Since ${\boldsymbol y}^*$ is a pure Nash equilibrium, we must have that no players can perform better response updates in ${\boldsymbol y}^*$, \rev{this proves Statement \rev{3)}}.

\rev{The proof of Statement \rev{4)} is similar to the proof of Statement \rev{2)}, and is hence omitted. \hfill $\Box$}



Next we prove the main Theorem \ref{poa} using Lemma \ref{usefullemma}.

Let ${\boldsymbol x}^*$ be a social optimum of the game. \rev{Let} ${\boldsymbol y}^*$ be a pure Nash equilibrium of the game which minimizes the social welfare \rev{among all pure Nash equilibria} (note that Theorem \ref{completeconvergence} implies that such a pure Nash equilibrium ${\boldsymbol y}^*$ exists for our game). Clearly \rev{the four statements in} Lemma \ref{usefullemma} hold in this scenario. Also Equation (\ref{poadfn}) gives
\begin{equation} \label{blueqn} \poa = \frac{\sum_{n=1} ^N U_n ( { \boldsymbol x }^*) }{ \sum_{n=1} ^N U_n ( { \boldsymbol y }^*)  }. \end{equation}
Now we shall prove statements (\ref{x3})-(\ref{ycond}) on by one:
\begin{equation} \label{x3} B({\boldsymbol x}^*) \in \{ 1,2,\edi{\ldots},N \}. \end{equation}
\begin{equation} \label{xfact} B({\boldsymbol x}^*) \leq C \max\{T_n^c: n \in \mathcal{N}, c \in \mathcal{C} \}   \end{equation}
\begin{equation} \label{y3}  B({\boldsymbol y}^*) \in \{ 1,2,\edi{\ldots},N \}. \end{equation}
\begin{equation} \label{ycond}  \textit{If } B({\boldsymbol y}^*) < N, \textit{ then } B({\boldsymbol y}^*) \geq C \min\{T_n^c: n \in \mathcal{N}, c \in \mathcal{C} \}  \end{equation}



Consider the strategy profile ${\boldsymbol z}^*$ where user $n=1$ uses channel $c=1$, and all the other users are dormant. User $n=1$ must be satisfied in ${\boldsymbol z}^*$ since $I^1({\boldsymbol z}^*)=1\leq T_1^1$, and so the social welfare of ${\boldsymbol z}^*$ is $\sum_{n =1}^N U_n({\boldsymbol {\boldsymbol z}^*}) =1$. Since ${\boldsymbol x}^*$ is a social optimum, its social welfare $\sum_{n =1}^N U_n({\boldsymbol {\boldsymbol x}^*})$ must be greater than or equal to that of ${\boldsymbol z}^*$, and so
\begin{equation} \label{p1} \sum_{n =1}^N U_n({\boldsymbol {\boldsymbol x}^*}) \geq \sum_{n =1}^N U_n({\boldsymbol z^*}) =1. \end{equation}
Now combining Statement \rev{2)} of Lemma \ref{usefullemma} with Inequality (\ref{p1}) gives $B({\boldsymbol x}^*)\geq 1$. Also clearly $B({\boldsymbol x}^*)$ is an integer less than or equal to $N$, \rev{hence we have proved} Statement (\ref{x3}).

Let $c' \in \{1,2,\edi{\ldots},C\}$ be one of the channels with the most users under ${\boldsymbol x}^*$ (i.e., $I^{c'} ({\boldsymbol x}^*) =\max\{ I^c ({\boldsymbol x}^*) : c \in \mathcal{C} \}$). Now Statement \rev{2)} from Lemma \ref{usefullemma} implies
\begin{equation} \label{p2} B({\boldsymbol x}^*) = \sum_{c=1}^C I^c ({\boldsymbol x}^*) \leq \sum_{c=1}^C \left( I^{c'} ({\boldsymbol x}^*) \right) = C I^{c'} ({\boldsymbol x}^*). \end{equation}
Now Statement (\ref{x3}) gives $1 \leq B({\boldsymbol x}^*)$, and combining this with Inequality (\ref{p2}) gives us that $1 \leq \rev{C} I^{c'} ({\boldsymbol x}^*)$. \rev{Since $I^{c'} ({\boldsymbol x}^*)$ is an integer we must also have  $1 \leq I^{c'} ({\boldsymbol x}^*)$.} It follows that there must be some user $n'$ of channel $c'$ under ${\boldsymbol x}^*$ (i.e., ${\boldsymbol x}^* _{n'} = c'$). Now from Statement \rev{1)} of Lemma \ref{usefullemma}, we have that $n'$ is satisfied with using $c'$ under ${\boldsymbol x}^*$, and so it follows that
\begin{equation} \label{p3} I^{c'} ({\boldsymbol x}^*) \leq T_{n'} ^{c'} \leq \max\{T_n^c: n \in \mathcal{N}, c \in \mathcal{C} \}, \end{equation}
Combining Inequality (\ref{p2}) and Inequality (\ref{p3}) yields
$B({\boldsymbol x}^*) \leq C I^{c'} ({\boldsymbol x}^*) \leq C \max\{T_n^c: n \in \mathcal{N}, c \in \mathcal{C} \},$ and so we have proved Statement (\ref{xfact}).

We can prove $B({\boldsymbol y}^*)\geq 1$ by contradiction. If $B({\boldsymbol y}^*)\geq 1$ were false, then we would have $B({\boldsymbol y}^*)= 0$, and no channel would have any \rev{active} users. However, in this case user $n=1$ could do a better response update by changing to channel $c=1$, because $T_n^c \geq 1$. \rev{This contradicts our assumption that ${\boldsymbol y}^*$ is a pure Nash equilibrium,  hence we must} have that $B({\boldsymbol y}^*)\geq 1$. Also \rev{it is clear that} $B({\boldsymbol y}^*)\leq N$, \rev{hence we have proved} Statement (\ref{y3}).

To prove \rev{Statement} (\ref{ycond}), suppose that  $B({\boldsymbol y}^*) < N$. This implies that there are users which are not satisfied under ${\boldsymbol y}^*$. Also Statement \rev{3) from Lemma \ref{usefullemma}} implies that every user which is not satisfied under ${\boldsymbol y}^*$ is dormant, and so it follows that there must be some user $n^*$ \rev{that} is dormant in ${\boldsymbol y}^*$. Since ${\boldsymbol y}^*$ is a pure Nash equilibrium, we have that player $n^*$ cannot do a better response by \rev{switching} to use a channel $c$. It follows that, for each channel $c \in \{1,2,\edi{\ldots},C\},$ we must have that
\begin{equation} \label{qq1} I^c ({\boldsymbol y}^*) \geq T_{n^*} ^c \geq \min\{T_n^c: n \in \mathcal{N}, c \in \mathcal{C} \}, \end{equation}
Now combining Statement \rev{4) from Lemma \ref{usefullemma}} with Inequality (\ref{qq1}), \rev{we have} $B({\boldsymbol y}^*) = \sum_{c=1}^C I^c({\boldsymbol y}^*) \geq \sum_{c=1}^C  \min\{T_n^c: n \in \mathcal{N}, c \in \mathcal{C} \} = C \min\{T_n^c: n \in \mathcal{N}, c \in \mathcal{C} \},$ \rev{which proves Statement (\ref{ycond})}.

Now we can prove \rev{Theorem \ref{poa}}. By taking Equation (\ref{blueqn}) and using Statements \rev{2)} and \rev{4)} from Lemma \ref{usefullemma}, one obtains $\poa = \frac{B({\boldsymbol x}^*)}{B({\boldsymbol y}^*)}$. Statement (\ref{x3}) gives $B({\boldsymbol x}^*) \leq N$, Statement (\ref{y3}) gives $B({\boldsymbol y}^*) \geq 1$, and so we must have $\poa \leq N$.

 \rev{Next consider two cases. In the first we have} $B({\boldsymbol x}^*) = B({\boldsymbol y}^*),$ and so we have $\poa =1$. Now since $1 \leq \frac{\max\{T_n^c: n \in \mathcal{N}, c \in \mathcal{C} \}}{\min\{T_n^c: n \in \mathcal{N}, c \in \mathcal{C} \}},$ Theorem \ref{poa} clearly holds in this case.
\rev{Consider the second case where $B({\boldsymbol x}^*) \neq B({\boldsymbol y}^*).$} In this case we must have $B({\boldsymbol x}^*) > B({\boldsymbol y}^*),$ because ${\boldsymbol x}^*$ is a social optimum. Moreover, Statement (\ref{x3}) \rev{implies} that $N \geq B({\boldsymbol x}^*)$ so $N> B({\boldsymbol y}^*)$. It follows from Statement (\ref{ycond}) that we must have
\begin{equation} \label{nearend} B({\boldsymbol y}^*) \geq C \min\{T_n^c: n \in \mathcal{N}, c \in \mathcal{C} \}.\end{equation}
Since $\poa = \frac{B({\boldsymbol x}^*)}{B({\boldsymbol y}^*)}$, we have that Inequality (\ref{nearend}) implies
\begin{equation} \label{nearnearend} \frac{B({\boldsymbol x}^*)}{\poa} \geq C \min\{T_n^c: n \in \mathcal{N}, c \in \mathcal{C} \}. \end{equation}
Now rearranging Inequality (\ref{nearnearend}), and combining with Inequality (\ref{xfact}) gives
$$\poa \leq \frac{C \max\{T_n^c: n \in \mathcal{N}, c \in \mathcal{C} \}}{C \min\{T_n^c: n \in \mathcal{N}, c \in \mathcal{C} \}}.$$
Cancelling the $C$s from this inequality, and combining \rev{it} with the inequality $\poa \leq N$ (which we have already established), \rev{we have} Inequality (\ref{poaTH}) in Theorem \ref{poa}. $\Box$
\subsection{Proof of Theorem \ref{homplayequiv}}\label{Proof5}
Let $B({\boldsymbol x}) = |\{ n \in \mathcal{N}: U_n ({\boldsymbol x}) = 1 \}|$ denote the number of satisfied users in a strategy profile ${\boldsymbol x}$. We will show that Statement $\rev{1)}$ implies Statement $\rev{2)}$, which in turn implies Statement $\rev{3)}$, which in turn implies Statement $\rev{1)}$.
\subsubsection{Statement \rev{1)}$\Rightarrow$ Statement \rev{2)}}
Suppose Statement $\rev{1)}$ holds, and ${\boldsymbol x}$ is a pure Nash equilibrium. Now Lemma \ref{usefullemma} implies that there are no suffering users in ${\boldsymbol x}$, and Lemma \ref{usefullemma} also implies that
 \begin{equation} \label{welfareeqn} B({\boldsymbol x})= \sum_{c=1} ^C I^c ({\boldsymbol x}) = \sum_{n=1}^N U_n({\boldsymbol x}).\end{equation}
 Since there are no suffering users under ${\boldsymbol x}$, we must have that $I^c ({\boldsymbol x}) \leq T^c$, for each $c \in \{1,2,\edi{\ldots}, C\}$. It follows that
  $B({\boldsymbol x})= \sum_{c=1} ^C I^c ({\boldsymbol x}) \leq \sum_{c=1} ^C T^c.$ Since we also have $B({\boldsymbol x}) \leq N$, it follows that
  \begin{equation} \label{twoineq} B({\boldsymbol x}) \leq \min\left\{ N, \sum_{c=1} ^C T^c \right\}. \end{equation}

\rev{Next consider two cases. In the first case with} $B({\boldsymbol x})=N,$ clearly Inequality (\ref{twoineq}) implies that $B({\boldsymbol x})=N \leq \sum_{c=1} ^C T^c$ and so ${\boldsymbol x}$ satisfies Statement $\rev{2)}$ of Theorem \ref{homplayequiv}.

  Now let us consider the \rev{second case where} $B({\boldsymbol x}) < N$. In this case there \rev{exists at least one} user $n^*$ that is not satisfied. We know that $n^*$ \rev{must be} dormant, since ${\boldsymbol x}$ contains no suffering users. Since ${\boldsymbol x}$ is a pure Nash equilibrium, we know that user $n^*$ cannot perform any best response updates. \rev{This implies that} $I^c({\boldsymbol x}) \geq T^c$, for each $c \in \{1,2,\edi{\ldots},C\}$. It follows that we must have
    \begin{equation} \label{bitmore} \sum_{c=1} ^C I^c ({\boldsymbol x}) \geq \sum_{c=1} ^C T^c.  \end{equation}
Combining Equation (\ref{welfareeqn}) and Inequality (\ref{twoineq}) gives us that
\begin{equation} \label{somethingextra} B({\boldsymbol x})= \sum_{c=1} ^C I^c ({\boldsymbol x}) \leq \sum_{c=1} ^C T^c, \end{equation}
 and combining Inequality (\ref{somethingextra}) with Inequality (\ref{bitmore}) yields
 \begin{equation} \label{isthistheend} B({\boldsymbol x}) = \sum_{c=1} ^C I^c ({\boldsymbol x}) = \sum_{c=1} ^C T^c. \end{equation}
 \rev{Since we have assumed} $B({\boldsymbol x}) < N$ \rev{in this second case}, we have $B({\boldsymbol x}) = \sum_{c=1} ^C T^c = \min\{ N, \sum_{c=1} ^C T^c \}.$ \rev{This shows that Statement \rev{1)} implies Statement \rev{2)}.}


\subsubsection{Statement \rev{2)}$\Rightarrow$ Statement \rev{3)}}
Now \rev{we assume} Statement \rev{2)} holds. If $B({\boldsymbol x})= N$, then ${\boldsymbol x}$ is clearly a social optimal and so Statement \rev{3)} follows in this case. Suppose instead that
\begin{equation} \label{thiscasee} B({\boldsymbol x}) = \sum_{c=1}^C T^c <N. \end{equation} Since ${\boldsymbol x}$ has no suffering users, we must have
\begin{equation} \label{longone} B({\boldsymbol x}) = |\{ n \in \mathcal{N} : x_n \neq 0\}| = \sum_{n=1}^N U_n ({\boldsymbol x}) = \sum_{c=1}^C I^c ({\boldsymbol x}) =  \sum_{c=1}^C T^c. \end{equation}
Let ${\boldsymbol z}$ be a social optimum of our game. \rev{From} Lemma \ref{usefullemma} \rev{we have}
\begin{equation} \label{sleepeqn} B({\boldsymbol z}) = \sum_{n=1}^N U_n ({\boldsymbol z}) = \sum_{c=1}^C I^c ({\boldsymbol z}). \end{equation} Since Lemma \ref{usefullemma} implies that ${ \boldsymbol z}$ holds no suffering users, we must have $I^c ( { \boldsymbol z}) \leq T^c$ for each $c \in \{1,2,\edi{\ldots},C\}$, and it follows that
\begin{equation} \label{kindamagic} \sum_{c=1}^C I^c ({\boldsymbol z}) \leq \sum_{c=1}^C T^c. \end{equation} Combining Equation (\ref{sleepeqn}), Inequality (\ref{kindamagic}), and Equation (\ref{longone}) yields
\begin{equation} \label{VVlongone}
\sum_{n=1}^N U_n ({\boldsymbol z}) =\sum_{c=1}^C I^c ({\boldsymbol z}) \leq \sum_{c=1}^C T^c = \sum_{n=1}^N U_n ({\boldsymbol x}). \end{equation} Inequality (\ref{VVlongone}) implies that the social welfare of ${\boldsymbol x}$ is no less than the social welfare of the social optimum ${\boldsymbol z}$. This implies that ${\boldsymbol x}$ is a social optimum, \rev{which proves Statement \rev{3)}}.

\subsubsection{Statement \rev{3)}$\Rightarrow$ Statement \rev{1)}}
%
%
Now \rev{we assume that} Statement \rev{3)} holds, and ${\boldsymbol x}$ is a social optimum. In this case, Lemma \ref{usefullemma} implies that there are no suffering users under ${\boldsymbol x}$. \rev{Next} we will show that ${\boldsymbol x}$ is a pure Nash equilibrium by contradiction.

Suppose ${\boldsymbol x}$ is not a pure Nash equilibrium. There must exist a player $n^* \in \mathcal{N}$ that can perform a better response update. \rev{This means that} $n^*$ must be dormant, because ${\boldsymbol x}$ contains no suffering users. It follows that $U_{n^*} ({\boldsymbol x}) = 0$, and there must exist some channel $c^* \neq 0$ such that $I^{c^*} ({\boldsymbol x}) < T^{c^*}$ (which $n^*$ can do a better response update by switching to). Let
$${\boldsymbol y} = (x_1,\edi{\ldots},x_{n^*-1},c^*,x_{n^*+1},\edi{\ldots},x_N)$$
be the strategy profile obtained by \rev{allowing} player $n^*$ \rev{to switch} to channel $c^*$. \rev{We shall have} $I^{c^*} ({\boldsymbol y}) = I^{c^*} ({\boldsymbol x}) +1  \leq  T^{c^*},$ so the users of \rev{channel} $c^*$ will still all be satisfied in ${\boldsymbol y}$. \rev{This implies that} $\sum_{n=1} ^N U_n({ \boldsymbol y}) = 1 + \sum_{n=1} ^N U_n({ \boldsymbol x}),$ which contradicts our assumption that ${\boldsymbol x}$ is a social optimum. \rev{This shows that Statement \rev{3)} implies Statement \rev{1)}}. $\Box$


\subsection{Brief sketch of the proof to Theorem \ref{algodworks}}\label{ProofSketch}

\rev{We order the users so that} $T_1 \geq T_2 \geq \edi{\ldots}. \geq T_N$. We use $\boldsymbol{x^n}$ \rev{to denote the strategy profile produced by the $n$th iteration of Algorithm \ref{d}}. Also $B(\boldsymbol{x})$ is the number of satisfied users in $\boldsymbol{x}$. We say a strategy profile $\boldsymbol{y}$ is \textbf{reachable} from strategy profile $\boldsymbol{x}$,
\rev{if for any
$x_p \neq 0$ (for a player $p \in \mathcal{N}$) we  have $y_p = x_p$. In other words, $\boldsymbol{y}$ is reachable from $\boldsymbol{x}$ if each player who is not dormant in $\boldsymbol{x}$ uses the same channel in $\boldsymbol{y}$ as it does in $\boldsymbol{x}$.} Let $\beta(\boldsymbol{x})$ denote the maximum value of $B(\boldsymbol{y})$ such that $\boldsymbol{y}$ is a natural strategy profile reachable from $\boldsymbol{x}$. Let $\mathcal{D}(\boldsymbol{x})$ denote the set of dormant users in the strategy profile $\boldsymbol{x}$.

 The key idea of the proof is to show that a social optimal is reachable from the strategy profile $\boldsymbol{x^n}$, $\forall n \in \{1,2,\edi{\ldots},N\}$ (here $\boldsymbol{x^n}$ is the strategy profile outputted by the $n$th iteration of Algorithm \ref{d}). This can be achieved by showing $\beta(\boldsymbol{x^0})=\beta(\boldsymbol{x^1})=\ldots=\beta(\boldsymbol{x^N})$. The reason is that $\beta(\boldsymbol{x^0})$ is the number of satisfied users at a social optimum (since all strategy profiles can be reached from $\boldsymbol{x^0}$).

Since a (natural) social optimum is reachable from $\boldsymbol{x^N}$ and since we can show\footnote{Our algorithm only stops altering the strategy profile when all users are satisfied, or when no more dormant users can be satisfied. It follows that no natural strategy profiles can be reached from $\boldsymbol{x^N}$ except $\boldsymbol{x^N}$.} that the only natural strategy profile that is reachable from $x^N$ is $x^N$ itself, we have that $x^N$ is a social optimum. By checking that users have no incentive to change, we can then show that the social optimum is also a pure Nash equilibrium.

\rev{In order to prove that a social optimal is always reachable from $\boldsymbol{x^n}$, we use induction to prove that $\boldsymbol{x^{n-1}}$ satisfies various conditions for each $n \in \{1,2,\edi{\ldots},N\}$.
In particular, we show that if a value $n \in \{1,2,\edi{\ldots},N\}$ is such that there exist
%
 a channel $c$ with the property that $I^c(\boldsymbol{x^{n-1}})<T_n$, then $\boldsymbol{x^{n-1}}$ satisfies the following conditions:}\com{see my change of last sentence is ok.}
 \comg{Basically ok, I just changed the tense a bit}

\begin{enumerate}
\item $\boldsymbol{x^{n-1}}$ is natural.
\item $\beta(\boldsymbol{x^{n-1}})=\beta(\boldsymbol{x^{0}}).$
\item $\mathcal{D}(\boldsymbol{x^{n-1}})=\{n,n+1,\edi{\ldots},N\}.$
\item $\{c\in\mathcal{C}:I^{c}(\boldsymbol{x^{n-1}})<T_{n}\}\neq\emptyset.$
\item \rev{Let $c^{*}=\min\{c\in\mathcal{C}:I^{c}(\boldsymbol{x^{n-1}})<T_{n}\}$. Then for each channel $c$, we have  {(i)} if $c<c^*$, then $I^{c}(\boldsymbol{x^{n-1}})\geq T_{n}$,  {(ii)} if $c=c^*$, then $I^{c}(\boldsymbol{x^{n-1}})<T_{n}$, and {(iii)} if $c>c^*$, then $I^{c}(\boldsymbol{x^{n-1}})=0$.}\com{check if statement 5) is still accurate after the revision.} \comg{its cool}
\end{enumerate}

\rev{We now provide the more details of the proof as follows.}

Let $B({\boldsymbol x}) = |\{ n \in \mathcal{N}: x_n \neq 0 \}|$ denote the number of satisfied players in strategy profile $x$.
We say a strategy profile $y$ is \textbf{reachable} from strategy profile $x$ when $x_n \neq 0$ implies $y_n = x_n$, for each $n \in \mathcal{N}$. In other words $y$ is reachable from $x$ when $x$ can be converted to $y$ by allocating real channels to dormant users.
Let $\beta({\boldsymbol x})$ denote the maximum number of satisfied users in a strategy profile reachable from $x$.

Algorithm 1 works by initializing all users with the off channel, and then having each player switch onto the most congested (and lowest indexed) channel they can benefit from using. Our proof of the validity of Algorithm 1 works by showing that, at any stage, a social optimal is reachable from the current profile considered. In particular, our algorithm initiates from the strategy profile $x^0$ were all users are `off'. Every strategy profile is reachable from this initial condition, and so $\beta(x^0)$ is equal to the maximum number of satisfied users in any strategy profile of the game. For brevity, se shall refer to social optima simply as `optima' or `optimal strategy profiles'.

The following lemma is the critical part of our proof of the validity of Algorithm 1, for it essentially asserts that $\beta(x^{n-1}) = \beta(x^{n})$ for each $n \in \{1,2,..,N\}$. In other words, as one iterates the algorithm, the $n$th profile generated, $x^n$, has just as beneficial strategy profiles that can be reached from it, as the $(n-1)$th profile $x^{n-1}$ had. This result can be used with induction to show that $\beta(x^0) = \beta(x^N)$. Moreover $\beta(x^N) = B({\boldsymbol x})$ is the number of satisfied users in the outputted strategy profile because once $x^N$ has been generated, we either have that all players have been allocated a real channel (in which case $x^N$ is the only strategy profile reachable from $x^N$), or all dormant users cannot benefit from using a real channel because their thresholds are less than or equal to the congestion level of each active channel (in which case every strategy profile reachable from $x^N$, other than $x^N$ has suffering users). We shall state and prove this critical lemma before we continue with our proof.

\textbf{Lemma 5.5}

Let $g$ be a QoS satisfaction game with $C > 1$ channels (which are homogenous) and $N$ players, with thresholds $T_1 \geq T_2 \geq ... \geq T_N$. Suppose $x$ satisfies the following conditions:
\begin{enumerate}
\item We have that $k({\boldsymbol x}) := \{n \in \mathcal{N} : x_n \neq 0 \}$ is non-empty, and $n^* \in \mathcal{N} : T_{n^*} = \max \{ T_n : n \in k({\boldsymbol x}) \}$ is a player with maximal threshold in $k({\boldsymbol x})$.
\item We have that $\exists F \in \{0,1,..,C-1\}$ such that for each $c \in \{1,2,..,C\}$ we have $c \leq F \Rightarrow I^c({\boldsymbol x})\geq T_{n^*}$ and $c > F \Rightarrow I^c({\boldsymbol x}) <T_{n^*}$ and $c>F+1 \Rightarrow I^c({\boldsymbol x}) =0$.
\item We have $\forall n, m \in \mathcal{N}$ that if $n \notin k({\boldsymbol x})$ and $m \in k({\boldsymbol x})$ then $T_n \geq T_m$.
\item We have that $x$ has no suffering users (i.e., $x$ is natural).
\end{enumerate}
Let $y$ denote the strategy profile obtained by taking $x$ and having player $n^*$ change their channel to $F+1$. Now $\beta(y) = \beta({\boldsymbol x})$.

\textbf{Proof of Lemma 5.5}

We shall construct a strategy profile $\Omega$ that is reachable from $x$ and such that $\Omega_{n^*} = F+1$ and $B(\Omega) = \beta({\boldsymbol x})$. Since $\Omega$ is reachable from $y$ we shall then have $\beta ({\boldsymbol x}) = B(\Omega) = \beta(y)$. We construct $\Omega$ by starting with a natural strategy profile $z$ that maximizes the number of benefitting users amongst those profiles reachable from $x$. Then we modify $z$ to make another strategy profile $w$. Then we modify $w$ to make $\Omega$.

Let $z:B(z) = \beta({\boldsymbol x})$ be a strategy profile that is reachable from $x$. Suppose this profile $z$ has the maximum number of satisfied users amongst all strategy profiles reachable from $x$. Also, suppose that $z$ has no suffering users.

Now note that $B(z) >B({\boldsymbol x})$, to see this note that points (1) and (2) above imply that, from $x$, user $n^*$ can beneficially start using channel $F+1$ without causing any other player to cease satisfied. It follows that $x$ there are strategy profiles (such as $z$) with more satisfied users, that are reachable from $x$.

Next we claim that there must exist some player $m \in k({\boldsymbol x})$ such that $z_m \in \{F+1,F+2,..,C\}$. To see this note that $B(z) >B({\boldsymbol x})$ implies that there is some $m \in k({\boldsymbol x}): z_m \neq 0$ and since $m$ must be satisfied in $z$, and $m \in k({\boldsymbol x})$ implies $T_m \leq T_{n^*} \leq I^c({\boldsymbol x}) \forall c \in \{1,2,..,F\}$ we must have $z_m \in \{F+1,F+2,..,C\}$.

If $z_{n^*} \neq 0$ then similarly we have $z_{n^*} \in \{F+1,F+2,..,C\}$, and in this case we let $w =z$. Now, alternatively suppose that $z_{n^*} =0$. In this case, let $w$ be the strategy profile obtained by taking $z$ and interchanging the strategies of $n^*$ and $m$. In other words, $\forall n \in \mathcal{N}$ we have $w_n = z_n$ if $n \notin \{n^*,m\}$, and $w_n = z_m$ if $n = n^*$ and $w_n = z_{n^*}$ if $n = m$.

Clearly $w$ is reachable from $x$. Also, note that $w$ (like $z$) has no suffering users. To see this, we just have to note that $w$ is just like $z$ except that we have replaced the real channel user $m$, with the user $n^*$, on the same channel. Now since $m \in k({\boldsymbol x})$ and $n^*$ has the maximum threshold of any user in $k({\boldsymbol x})$ we must have that $T_{n^*} \geq T_m$. Now, since $m$ was satisfied in $z$ it follows that, when we replace $m$ with player $n^*$ (using the same channel), we shall have that $n^*$ is satisfied in the resulting strategy profile $w$. The reason is that $n^*$ incurs exactly the same congestion level in $w$ as $m$ incurred in $z$, and $T_{n^*} \geq T_m$. Also $B(w) = B(z)$, since the operation we use to obtain $w$ from $z$ preserves the number of users of real channels.

So now we have that $w$ is reachable from $x$ and $B(w) = B(z) = \beta({\boldsymbol x})$. If $w_{n^*} = F+1$ then let $\Omega = w$, and we are done. Now instead suppose $w_{n^*} \neq F+1$. We shall describe how to construct $\Omega$ in this case.

If $I^{F+1}(w) \geq I^{w_{n^*}}(w)$ then we construct $\Omega$ by taking $w$ and swapping around the channels of $n^*$ and the member $m'$ of
$$R:=\{n \in \mathcal{N} : w_n = F+1 \neq x_n \} \subseteq k({\boldsymbol x})$$
 with the highest threshold. In other words, for each $n \in \mathcal{N}$ we have $n \notin \{n^*, m'\}$ implies $\Omega_n = w_n$ and $n = n^*$ implies $\Omega _n = w_{m'}$ and $n = m'$ implies $\Omega _n = w_{n^*}$.

The player $m' \in R$ which changes their channel from $F+1$ to $w_{n^*}$ under this operation will not stop being satisfied, since they end up (in $\Omega$) using a channel $w_{n^*}$ that is no more congested than the channel $F+1$ they were using in $w$. Also, player $n^*$, who changes their channel from $w^*$ to $F+1$, will not stop being satisfied since they end up (in $\Omega$) with the same congestion level as $m'$ had in $w$, but they have a threshold that is greater than or equal to that of $m'$. It follows that $n^*$ and $m'$ (and each other user of a real channel) will be satisfied in $w$. This shows $\Omega$ is natural, and $B(\Omega) = B(w)$. Also, $\Omega$ is clearly reachable from $x$ since it can be obtained by taking $z$ and altering the actions of some users from $k({\boldsymbol x})$ who were ``off'' in $x$.

Now let us consider how to define $\Omega$ in the final case, where $w_{n^*} \neq F+1$ and $I^{F+1} (w) < I^{w_{n^*}}(w)$. In this case we get $\Omega$ by having the $I^{w_{n^*}} (w) - I^{F+1} ({\boldsymbol x})$ players with the highest thresholds that are using $w_{n^*}$ under $w$, change their channels to $F+1$, whilst (simultaneously) each player from $R$ changes their channel from $F+1$ to $w_{n^*}$. Let us be more precise. Let us name the users of $w_{n^*}$ under $w$ by writing $\{ n \in \mathcal{N} : w_n = w_{n^*} \} = \{e_1, e_2, .., e_M\}$. Here we have given the players names $e_i$ in such a way that $e_1 = n^*$ and $T_{e_1} \geq T_{e_2} \geq .. \geq T_{e_M}$. Now in this case, $\Omega$ is defined such that $\forall n \in \mathcal{N}$ we have $n \notin R \cup \{e_1,e_2,..,e_{I^{w_{n^*}} (w) - I^{F+1} ({\boldsymbol x})}\}$ implies $\Omega _n = w_n$, and $n \in R$ implies $\Omega_n = w_{n^*}$, and $n \in \{e_1,e_2,..,e_{I^{w_{n^*}} (w) - I^{F+1} ({\boldsymbol x})}\}$ implies $\Omega_n = F+1$.

Now clearly $\Omega$ is reachable from $x$ since it is obtained but taking $w$ and only altering the channels of users from $R, \{ n \in \mathcal{N} : w_n = w_{n^*} \} \subseteq k({\boldsymbol x})$. Now we will show that $\Omega$ has no suffering users.

Firstly, the players in the set $R$ that change their channels from $F+1$ to $w_{n^*}$ will still be satisfied in $\Omega$. The reason is that the congestion level these players experience in $\Omega$ will be $I^{F+1}({\boldsymbol x}) + |R|$, which is the same as the congestion level that they incurred in $w$.

Players sticking on $w_{n^*}$ in $w$ and $\Omega$ experience a congestion level $I^{w_{n^*}}(\Omega) = I^{F+1}({\boldsymbol x}) + |R| = I^{F+1}(w) < I^{w_{n^*}}(w)$ in $\Omega$ that is less than the congestion level that they incurred in $w$, so these players will still be satisfied in $\Omega$.

Players in the set $\{e_1,e_2,..,e_{I^{w_{n^*}} (w) - I^{F+1} ({\boldsymbol x})}\}$ that change their channels from $w_{n^*}$ to $F+1$ will still be satisfied in $\Omega$, since the congestion level  $I^{F+1} (\Omega) = I^{T+1} ({\boldsymbol x}) + I^{w_{n^*}}(w) -  I^{T+1} ({\boldsymbol x})  = I^{w_{n^*}}(w)$ that these users experience in $\Omega$ will be the same as the congestion levels they experienced in $w$.

Also, the players $\{ n \in \mathcal{N} : x_n = F+1 \} \subseteq \mathcal{N} - k({\boldsymbol x})$ which stick upon channel $F+1$ in $w$ and $\Omega$ each have thresholds greater than or equal to each player in $k({\boldsymbol x})$ (according to point 3) and so, since the players $\{e_1,e_2,..,e_{I^{w_{n^*}} (w) - I^{F+1} ({\boldsymbol x})}\} \subseteq k({\boldsymbol x})$ are satisfied upon channel $F+1$ in $\Omega$, it follows that the players $\{ n \in \mathcal{N} : x_n = F+1 \}$ are also satisfied in $\Omega$.

So we have shown that each user of $w_{n^*}$ or $F+1$ is satisfied in $\Omega$. Now since $w_{n^*}$ or $F+1$ are the only channels that have different user sets in $\Omega$ and $w$, we have that $B(\Omega) = B(w) = B(z) = \beta ({\boldsymbol x})$ and $\Omega$ has no suffering users. Also note that $\Omega$ is reachable from $x$. Also, since $\Omega_{n^*} =T+1$, we have that $\Omega$ is reachable from $y$. From this it follows that $\beta(y) = B(\Omega) = \beta({\boldsymbol x})$.

$\Box$

Now we have proved Lemma 5.5, we shall continue our proof of Theorem 5.


Algorithm 1 runs so that we begin with all players using the zero-channel and then we update the players in order of descending threshold. $\forall n \in \{1,2,.., N\}$, we obtain the strategy profile $x^n$ during the $n$th iteration of our algorithm. Here $x^n$ is obtained from $x^{n-1}$ by updating player $n$ (who has the $n$th highest threshold). To update player $n$ we check if there are any channels available, that have low enough congestion levels to satisfy user $n$. If such channels exist then player $n$ starts using the one $c^*$ with the lowest index, and this action produces the strategy profile $x^n$. Otherwise, if there is no channel with a low enough congestion level for player $n$ to benefit from using, then we say that $x^n$ is ``full''. In this case we will have that $x^n = x^{n+1} = .. x^N$ because every subsequent update will involve getting a user $n' >n$, with a threshold $T_{n'} \leq T_n$ and checking if there is a channel with a congestion level below the threshold $T_{n'}$ (i.e., stage 3 of the algorithm), and there will be no channel with congestion level below $T_{n'}$, because there was no channel with congestion level below $T_{n}$. So it follows that once our algorithm hits a full $x^n$, it will output $x^N = x^n$ eventually.

We say a state $x$ has property $(F,n^*)$ when the following conditions hold:
\begin{enumerate}
\item We have that $k({\boldsymbol x}) := \{n \in \mathcal{N} : x_n \neq 0 \}$ is non-empty, and $n^* \in \mathcal{N} : T_{n^*} = \max \{ T_n : n \in k({\boldsymbol x}) \}$ is a player with maximal threshold in $k({\boldsymbol x})$.
\item We have that $\exists F \in \{0,1,..,C-1\}$ such that for each $c \in \{1,2,..,C\}$ we have $c \leq F \Rightarrow I^c({\boldsymbol x})\geq T_{n^*}$ and $c > F \Rightarrow I^c({\boldsymbol x}) <T_{n^*}$ and $c>F+1 \Rightarrow I^c({\boldsymbol x}) =0$.
\item We have $\forall n, m \in \mathcal{N}$ that if $n \notin k({\boldsymbol x})$ and $m \in k({\boldsymbol x})$ then $T_n \geq T_m$.
\item We have that $x$ has no suffering users.
\end{enumerate}

Clearly $x^0$ has property $(0,1)$ since $c >0 \Rightarrow I^c (x^0) = 0 < T_1$ and $k({\boldsymbol x}) = \{1,2,..,N\}$.

Now we claim (**) that $\forall n \in \{1,2,..,N-1\}$ that if $x^{n-1}$ has property $(F,n)$ and $x^{n-1}$ is not full then the strategy profile $x^n$ present upon the next iteration of our algorithm will have property $(F',n+1)$, for some $F'$.

To see this note that if $x^{n-1}$ has property $(F,n)$ and $x^{n-1}$ is not full then $c^* = \min \{c \in \{1,2,..,C\} : I^{c}(x^{n-1}) <T_n\} = F+1$ and $x^n$ is the strategy profile obtained by having player $n$ start using channel $F+1$.

This move will not cause any of the previous users of channel $F+1$ to cease satisfied (since all their thresholds are at least as large as $n$'s threshold). Also, $n$ will clearly be satisfied in $x^n$. It follows that $x^n$ has no suffering users. Also, it follows that $k(x^n) = k(x^{n-1}) - \{n \}= \{n+1,n+2,..,N\}$, where $n+1$ is the player with the maximum threshold in $x^n$ that is not satisfied.

Now either $I^{c^*}(x^{n-1}) +1 < T_{n+1}$ in which case $x^n$ has property $(F,n+1)$, since $F = c^*$ is still the channel that $n+1$ is destined to switch to, or $I^{c^*}(x^{n-1}) +1 \geq  T_{n+1}$, in which case $x^n$ has property $(F+1,n+1)$, since $F+1 = c^*+1$ is the channel that the $n+1$ is destined to switch to.

Now since the initial condition $x^0 = (0,0,..,0)$ has property $(F,n^*)$ for some $F$ and $n^*$, we can use induction to show that every time $x^n \neq x^{n-1}$ (i.e., every time the system is not full) we have that $x^n$ has property $(A,B)$ for some $(A,B)$.

If $x^i$ has the property with respect to $(A_i,B_i)$ then Lemma 5.5 implies that $x^{i+1}$ has the property that $\beta(x^i) = \beta(x^{i+1})$.

Now if the system never gets full then we have that $x^0, x^1,.., x^{N-1}$ each $x^i$ has the property, for some pair $(A_i,B_i)$. It follows that
$\beta(x^0) = \beta(x^1)=.. \beta(x^{N-1}) = \beta(x^{N}) = B(x^{N})$, where $x^N$ is the output. The reason $\beta(x^N) = B(x^N)$ is because every user is satisfied in $x^N$ in this case, so $x^N$ is the only strategy profile reachable from $x^N$.

Similarly, if the system gets full up on time step $j$ then we have, for each $i \in \{1 , 2 ,.., j\}$ that $x^i$ has the property, for some pair $(A_i,B_i)$. It follows that $\beta(x^j) = \beta(x^0)$. Moreover since $x^j$ is full, we have $x^j = x^{j+1} = .. = x^N$, and so $\beta(x^0) = \beta (x^N)$. Also, since $x^N$ is full, no more users can be made to benefit from $x^N$ and so $\beta(x^N) = B(x^N)$.

And so we have shown that the maximum number of satisfied users that can be obtained in any strategy profile, which is $\beta (x^0)$, is equal to the number of satisfied users in the strategy profile $x^N$ outputted by our algorithm. This shows that the output $x^N$ is an optimum strategy profile of the game.

Now we just have to show that $x^N$ is a pure Nash equilibrium. To see this note that if a user in $x^N$ did want to switch channels, then they would want to benefit. This would only be possible if the system has reached a full state previously (since if the algorithm never hits a full state then its output is optimal, and unsatisfied users do not exist). So there must have come some earlier time when the system became full. From this point on, we have that each channel has an congestion level too high for any user of an off channel to benefit from switching to them, and so in fact it is impossible for the final state $x^N$ to have any users which are not satisfied that can increase their utilities by increasing channels.

Individual executions of stages $1,2,3,4,5,6,7$ and $8$ of the algorithm can be performed in $O(N),O(1),O(C),O(C),O(N),O(1),O(1)$ and $O(1)$ time respectively. Only stages $3,4,5,6$ and $7$ are repeated. Each of these stages can be performed once in $O(CN)$ time. These procedure where each of these stages are performed once is repeated $N$ time. The complete procedure/loop (that is initiated on stage 2) this takes $O(CN^2)$ time. This $O(CN^2)$ term dominates the execution times of all other stages outside this loop, and so the total run time of the system is $O(CN^2)$.

\subsection{Proof of Theorem \ref{xuopt}}\label{Proof3}

Let $\boldsymbol{x}$ be the strategy profile where $x_n=1+(n \mod \rev{C})$, $\forall n \in \mathcal N$, $\forall c \in \mathcal C = \{1,2,\edi{\ldots},C\}$ (i.e., where the users are spaced as evenly across the channels as possible). For such a strategy profile, we have $I^c({\boldsymbol x}) \leq \left\lceil \frac{N}{C} \right\rceil \leq T_n,  \forall c \in \mathcal{C},$ and so every player is satisfied. $\Box$

\subsection{Proof of Theorem \ref{graphical}}\label{Proof4}

Let us define the function $\Phi$ (which maps strategy profiles to real numbers) such that for each strategy profile $\boldsymbol{x}$ we have $\Phi({\boldsymbol x}) = \left( \sum_{n \in \mathcal{N}: x_n \neq 0} T_n ^{x_n} \right) - \sum_{c = 1} ^C \left( |\{ \{n, m \} \in \rev{\mathcal{E}} : x_n = x_m = c \}| + \frac{ |\{ n \in \mathcal{N}: x_n = c \}|}{2} \right).$
Here $|\{ \{n, m \} \in \rev{\mathcal{E}} : x_n = x_m = c \}|$ is the number of edges linking players using channel $c$, and $|\{ n \in \mathcal{N}: x_n = c \}|$ is the number of players using channel $c$. In other words, $\Phi({\boldsymbol x})$ is equal to [the sum of the thresholds which the non-dormant users associate with their channels] minus [the number of edges linking users of the same channel] minus [half the number of non-dormant users].

Suppose player $n'$ does a better response update by changing their strategy from $c' \in \{0,1,\edi{\ldots},C\}$ to $d' \in \{0,1,\edi{\ldots},C\}$, and this has the effect of changing the strategy profile from $\boldsymbol{x}$ to $\boldsymbol{y}=(x_1,\edi{\ldots},x_{n'-1},d',x_{n'+1},\edi{\ldots},x_N)$.

\rev{Next} we will show $\Phi (\boldsymbol{y}) \geq \Phi ({\boldsymbol x}) + \frac{1}{2}$ in each of the three possible cases:
\begin{enumerate}
\item $c'=0$, $d' \neq 0$ (i.e., when $n'$ stops being dormant).
\item $c' \neq 0$, $d' = 0$ (i.e., when $n'$ becomes dormant).
\item $c' \neq 0$, $d' \neq 0$ (i.e., when $n'$ switches from one channel to another).
\end{enumerate}

In case \rev{1)}, where $c'=0$, $d' \neq 0$, we have $\Phi (\boldsymbol{y}) = \Phi ({\boldsymbol x}) + T_{n'}^{d'} - I_{n'}^{d'} ({\boldsymbol x}) - \frac{1}{2},$ because \rev{the action where player $n'$ switches to channel $d'$} increases the number of edges linking users of $d'$ by $I_{n'}^{d'} ({\boldsymbol x})$ and increases the number of players using resource $d'$ by $1$.
Also, since our move is a better response update, we have $U_{n'}({\boldsymbol x}) = 0$ and $U_{n'}(\boldsymbol{y}) = 1$, and so $T_{n'}^{d'} \geq I_{n'}^{d'} (\boldsymbol{y}) = I_{n'}^{d'} ({\boldsymbol x}) + 1$.
It follows that $\Phi (\boldsymbol{y}) - \Phi ({\boldsymbol x}) = T_{n'}^{d'} - I_{n'}^{d'} ({\boldsymbol x}) - \frac{1}{2} \geq \frac{1}{2}$.

In case \rev{2)}, where $c' \neq 0$, $d' = 0$ we have $\Phi (\boldsymbol{y}) = \Phi ({\boldsymbol x}) - T_{n'}^{c'} + I_{n'}^{c'} ({\boldsymbol x}) -1 + \frac{1}{2},$ because \rev{the action where player $n'$ leaves channel $c'$} decreases the number of edges linking users of $c'$ by $I_{n'}^{c'} ({\boldsymbol x})-1$ and decreases the number of users of resource $c'$ by $1$.
Also, since our move is a better response update, we have $U_{n'}({\boldsymbol x}) = -1$ and $U_{n'}(\boldsymbol{y}) = 0$. \rev{It follows that} $T_{n'}^{c'} < I_{n'}^{c'} ({\boldsymbol x})$, and since $T_{n'}^{c'}$ and $I_{n'}^{c'} ({\boldsymbol x})$ are both integers, this implies $T_{n'}^{c'} \leq I_{n'}^{c'} ({\boldsymbol x})-1$. It follows that $\Phi (\boldsymbol{y}) - \Phi ({\boldsymbol x}) =  I_{n'}^{c'} ({\boldsymbol x}) -1 - T_{n'}^{n'} + \frac{1}{2} \geq \frac{1}{2}$.

In case \rev{3)}, where $c' \neq 0$, $d' \neq 0,$ we have $\Phi (\boldsymbol{y}) = \Phi ({\boldsymbol x}) + T_{n'}^{d'} - I_{n'}^{d'} ({\boldsymbol x}) - T_{n'}^{c'} + I_{n'}^{c'} ({\boldsymbol x}) -1,$ because \rev{the action where player $n'$ switches from channel $c'$ to channel $d'$} increases the number of edges linking users of $d'$ by $I_{n'}^{d'} ({\boldsymbol x})$ and decreases the number of edges linking users of $c'$ by $I_{n'}^{c'} ({\boldsymbol x})-1$.
Also, since our move is a better response update, we have $U_{n'}({\boldsymbol x}) = -1$ and $U_{n'}(\boldsymbol{y}) = 1$, and so $T_{n'}^{d'} \geq I_{n'}^{d'} (\boldsymbol{y}) = I_{n'}^{d'} ({\boldsymbol x}) + 1$ and $T_{n'}^{c'} \leq I_{n'}^{c'} ({\boldsymbol x})-1$. It follows that $\Phi (\boldsymbol{y}) - \Phi ({\boldsymbol x}) \geq 1$.


Without loss of generality, we can suppose that $-1 \leq T_n^c \leq N+1$, $\forall n \in \mathcal{N}, \forall c \in \{1,2,\edi{\ldots}C\}$, since thresholds less than $-1$ induce the same kind of behavior as thresholds equal to $-1$ (i.e., they can never be satisfied) and thresholds greater than $N+1$ induce the same kind of behavior as thresholds equal to $N+1$ (i.e., they are always satisfied). For any strategy profile $\boldsymbol{x}$ we have $(-1)N \leq \left( \sum_{n \in \mathcal{N}: x_n \neq 0} T_n ^{x_n} \right) \leq N(N+1)$. It is also true that $0 \leq  \sum_{c = 1} ^C |\{ \{n, m \} \in \rev{\mathcal{E}} : x_n = x_m = c \}| \leq \frac{N(N-1)}{2}$ and $0 \leq \sum_{c = 1} ^C  \frac{ |\{ n \in \mathcal{N}: x_n = c \}|}{2} \leq \frac{N}{2}$. From these inequalities, it follows that $-N \leq \Phi({\boldsymbol x}) \leq N(N+1) + \frac{CN(N-1)}{2} + \frac{CN}{2} = N + \frac{3N^2}{2}.$

When we start to evolve our system, the value of $\Phi$ for the initial strategy profile cannot be less than $-N$. Also, the value of $\Phi$ will increase by at least $\frac{1}{2}$ with every better response update. Now suppose we have performed $t$ better response updates (i.e., we have \rev{run} the system for $t$ time slots) and arrived at strategy profile $\boldsymbol{y}$. We must have $ -N + \frac{t}{2} \leq \Phi(\boldsymbol{x}) + \frac{t}{2} \leq \Phi(\boldsymbol{y}) \leq N + \frac{3N^2}{2},$ because the value of $\Phi$ increases by at least $\frac{1}{2}$ on each time step. This implies $t \leq 4N + 3(N)^2.$

So \rev{far} we have shown that it is \emph{impossible} to run the system (with asynchronous better response updates) for more than $t=4N + 3(N)^2$ time slots. This implies that when we evolve the system under asynchronous better response updates, we \emph{must} reach a strategy profile $\boldsymbol{z}$ from which no further better response updates can be performed, within $4N + 3(N)^2$ time slots. Such a strategy profile $\boldsymbol{z}$ must be a pure Nash equilibrium by definition. $\Box$

\end{appendix}


\bibliographystyle{ieeetran}
\bibliography{QoS3}

\end{document}